\documentclass[usenatbib]{mnras}
\usepackage{newtxtext,newtxmath}
\usepackage[T1]{fontenc}

\usepackage{graphicx}	
\graphicspath{{figures/}}

\usepackage{amsmath}	
\usepackage{physics}
\usepackage{hyperref}   
\usepackage{booktabs}   
\usepackage{xurl}
\usepackage[normalem]{ulem}  
\usepackage{soul} 

\usepackage[acronym]{glossaries}
\usepackage{siunitx}

\newacronym{dm}{DM}{dark matter}
\newacronym{cdm}{CDM}{cold dark matter}
\newacronym{wdm}{WDM}{warm dark matter}
\newacronym{hmf}{HMF}{halo mass function}
\newacronym{los}{LOS}{line-of-sight}
\newacronym{mnre}{MNRE}{marginal neural ratio estimation}
\newacronym{tmnre}{TMNRE}{truncated marginal neural ratio estimation}
\newacronym{sple}{SPLE}{singular power-law ellipsoid}
\newacronym{ps}{PS}{power spectrum}
\newacronym{hst}{HST}{Hubble Space Telescope}
\newacronym{nfw}{NFW}{Navarro-Frenk-White}
\newacronym{tnfw}{tNFW}{truncated \gls*{nfw}}
\newacronym{psf}{PSF}{point-spread function}
\newacronym{snr}{SNR}{signal-to-noise}
\newacronym{ts}{TS}{test statistic}
\newacronym{mlr}{MLR}{maximum likelihood ratio}
\newacronym{mcmc}{MCMC}{Markov chain Monte-Carlo} 
\newacronym{cnn}{CNN}{convolutional neural network}
\newacronym{mlp}{MLP}{multi-layers perceptron}
\newcommand{\swyft}{\textit{swyft} }

\DeclareSIUnit \Mpc {Mpc}
\DeclareSIUnit \arcsec {arcsec}
\DeclareSIUnit \solmass {M_{\odot}}
\newcommand{\keV}{\kiloElectronvolt}

\newcommand{\Uniform}{\mathcal{U}}

\newcommand{\mhm}{\mathrm{M_{hm}}}
\newcommand{\bx}{\boldsymbol{x}}
\newcommand{\msub}{\boldsymbol{m}_{200, \mathrm{sub}}}
\newcommand{\mlos}{\boldsymbol{m}_{200, \mathrm{los}}}
\newcommand{\psub}{\vec{\boldsymbol{p}}_\mathrm{sub}}
\newcommand{\plos}{\vec{\boldsymbol{p}}_\mathrm{los}}
\newcommand{\zlos}{\boldsymbol{z}_\mathrm{los}}

\title[DM cutoff mass with TMNRE]{Estimating the warm dark matter mass from strong lensing images with truncated marginal neural ratio estimation}

\author[N. Anau Montel et al]{
Noemi Anau Montel$^{1}$\thanks{Email: n.anaumontel@uva.nl},
Adam Coogan$^{1,2,3}$\thanks{Email: adam.coogan@umontreal.ca},
Camila Correa$^{1}$,
Konstantin Karchev$^{1,4}$,
Christoph Weniger$^{1}$\thanks{Email: c.weniger@uva.nl}
\\
$^{1}$GRAPPA (Gravitation Astroparticle Physics Amsterdam), University of Amsterdam, Science Park 904, 1098 XH Amsterdam, The Netherlands\\
$^{2}$Département de Physique, Université de Montréal, 1375 Avenue Thérèse-Lavoie-Roux, Montréal, QC H2V 0B3, Canada \\
$^{3}$Mila – Quebec AI Institute, 6666 St-Urbain, 200, Montreal, QC, H2S 3H1 \\
$^{4}$SISSA (Scuola Internazionale Superiore di Studi Avanzati), via Bonomea 265, I-34136 Trieste, Italy.
}

\date{Accepted XXX. Received YYY; in original form ZZZ}

\pubyear{2022}

\begin{document}
\label{firstpage}
\pagerange{\pageref{firstpage}--\pageref{lastpage}}
\maketitle

\begin{abstract}
Precision analysis of galaxy-galaxy strong gravitational lensing images provides a unique way of characterizing small-scale dark matter halos, and could allow us to uncover the fundamental properties of dark matter's constituents. Recently, gravitational imaging techniques made it possible to detect a few heavy subhalos. However, gravitational lenses contain numerous subhalos and line-of-sight halos, whose subtle imprint is extremely difficult to detect individually. Existing methods for marginalizing over this large population of sub-threshold perturbers to infer population-level parameters are typically computationally expensive, or require compressing observations into hand-crafted summary statistics, such as a power spectrum of residuals. Here, we present the first analysis pipeline to combine parametric lensing models and a recently-developed neural simulation-based inference technique called truncated marginal neural ratio estimation (TMNRE) to constrain the warm dark matter halo mass function cutoff scale directly from multiple lensing images. Through a proof-of-concept application to simulated data, we show that our approach enables empirically testable inference of the dark matter cutoff mass through marginalization over a large population of realistic perturbers that would be undetectable on their own, and over lens and source parameters uncertainties. To obtain our results, we combine the signal contained in a set of images with Hubble Space Telescope resolution. Our results suggest that TMNRE can be a powerful approach to put tight constraints on the mass of warm dark matter in the multi-keV regime, which will be relevant both for existing lensing data and in the large sample of lenses that will be delivered by near-future telescopes.

\bigskip
\end{abstract}


\makeatletter
\@thanks
\makeatother


\section{Introduction}
\label{sec:intro}

Over the past several decades, numerous astrophysical probes including rotational curves of spiral galaxies \citep{Rubin_1980}, galaxy-cluster dynamics \citep{Zwicky_1933}, cosmic microwave background \citep{Planck_2015}, gravitational lensing observations \citep{Taylor_1998}, have established \gls*{dm} as one of the major components of the Universe. However, up to the present time, the fundamental nature of \gls*{dm} is still an unresolved puzzle. 
For many years, the \gls*{cdm} paradigm \citep{Peebles_1982} has been able to accurately reproduce vastly disparate large-scale observations across all epochs. In this model, \gls*{dm} is massive, neutral, non-relativistic, and collisionless. The main prediction of the \gls*{cdm} paradigm is that structure formation is due to a hierarchical clustering process, guided by gravitational instability of \gls*{dm} density perturbations, originated from quantum fluctuations during inflation. 

Despite providing a stunning description of the observed distribution of matter on large scales ($>\order{\si{\Mpc}}$), the agreement between \gls*{cdm} predictions and observations at galactic and sub-galactic scales has been less clear. One of the most well-known small-scales discrepancies of \gls*{cdm} is the missing satellites problem \citep{Moore_1999}. Numerical \gls*{cdm} simulations predict that a large population of \gls*{dm} subhalos, spanning a wide range of masses, should be orbiting around all main \gls*{dm} halos. However, we have observed a lot fewer small galaxies in the Local Group than the predicted subhalos below $10^9\ \si{\solmass}$ \citep{Klypin_1999}. 

Solutions to this tension include the impact of baryonic processes or alternative \gls*{dm} physics. 
Baryonic processes from supernovae feedback and reionization processes suppress star formation in low-mass galaxies \citep{Bullock_2010}. As a result, most \gls*{dm} subhalos would not contain sufficiently bright galaxies and thus are more difficult to detect.
The other approach requires an alteration of \gls*{dm} particle physics, such that large-scale predictions remain unaffected, but the number of small-scale substructures is suppressed. One of the alternative models that has been proposed is \gls*{wdm} \citep{Colin_2000, Lovell_2014}. Moreover, its main particle candidates, sterile neutrinos \citep{Boyarsky_2019} and gravitinos \citep{Bond_1982}, are well-motivated from a particle physics perspective. In \gls*{wdm} models \gls*{dm} particles have non-negligible thermal velocities that allow them to free-stream out of density perturbations, effectively preventing small-scale structure formation. The scale at which this happens depends on model parameters and is parametrised by the half-mode mass $\mhm$ in the \gls*{hmf}. 
Therefore, one of the viable way to discriminate between \gls*{cdm} and alternative \gls*{dm} models is to constrain the low-mass end of the \gls*{hmf} by probing small-scale \gls*{dm} halos which are completely devoid of stars and truly \emph{dark}, whose only signature is then gravitational. 

\paragraph*{Strong lensing images analysis.} In strong gravitational lensing, the gravitational field of a mass distribution acts as a lens by distorting and magnifying the light flux coming from a background source \citep{Kochanek_2004}. This effect is sensitive only to how matter is distributed, regardless of its physical nature (baryonic/DM). Hence, it provides a direct way of probing the distribution of \gls*{dm} at small scales, by means of the distortions to the images due to substructures on top of the main lens mass distribution. Therefore, gravitational lensing provides a pristine probe of small-scale structures and can in principle distinguish between \gls*{cdm} and \gls*{wdm} scenarios.

Various different methods have been suggested to analyse the effects of small-scale structures on lensing images \citep{Drlica-Wagner_2019}. These methods usually target two different types of lensing systems that differ in the lensed source: extended background galaxies that get lensed into extended arcs or complete Einstein rings, and almost point-like quasars that get lensed into multiple point-like projections. 

Quasar lensing was first used in \cite{Mao_1998} to constrain the amount of \gls*{dm} substructures by analysing the deviations in the relative fluxes of the multiple source projections from a smooth lens model. Later, \cite{Dalal_2002} derived a statistical constraint on the substructure fraction in the lensing galaxies using a small sample of seven lensed quasars. \cite{Nierenberg_2014} showed that flux-ratio anomalies can also be used to detect individual low-mass subhalos. Several studies derived upper limits on the half-mode mass $\mhm$ \citep{Nierenberg_2017, Gilman_2018}, also including perturbations due to \gls*{los} halos \citep{Gilman_2019a, Gilman_2019b}. Further investigations \citep{Hsueh_2016, Hsueh_2017, Hsueh_2019} pointed out the importance of correctly modeling baryonic structure in the main lens, in order to avoid systematic errors while constraining \gls*{dm} substructure abundance with flux-ratio anomalies. 

On the other hand, in strong-lens systems where the background source is a galaxy, massive substructures can leave a signature in the form of percent-level variations in the shape of the predicted lensed light based on a smooth lens model. The gravitational imaging technique, which models these distortions, was first introduced in \cite{Koopmans_2005} and further developed in \cite{Vegetti_2009a, Vegetti_2009b}. Its application to real data has lead to several detections of individual heavy ($>10^8 \si{\solmass}$) subhalos \citep{Vegetti_2010a, Vegetti_2010b, Vegetti_2012, Hezaveh_2016a}. Moreover, samples of gravitational lens systems have been analyzed in \cite{Vegetti_2014, Vegetti_2018}, and, including \gls*{los} halos modeling, in \cite{Ritondale_2019}, in order to derive constraints on the \gls*{hmf} using detections and non-detections of individual substructures.

A population of low-mass halos can collectively cause perturbations to images that can be detected statistically in order to constrain the \gls*{hmf}. 
In reality, constraining collective substructure properties from gravitational lensing images is an extremely difficult problem. In fact, inferring marginal posteriors for the \gls*{hmf} cutoff requires marginalizing over all source, lens, and substructures parameters to get the marginal likelihood for the population-level parameter of interest, thus involving a time-consuming exploration of a very high-dimensional parameter space for complex realistic models. Therefore, \gls*{mcmc} or nested sampling methods would imply an intractable sampling from the high-dimensional joint posterior. 

To partially overcome traditional likelihood-based methods' challenges, \citet{Brewer_2015} and \citet{Daylan_2018} performed inference on subhalos using a likelihood-based method called transdimensional Bayesian inference. This approach uses transdimensional \gls*{mcmc} sampling over the union of different models, with different numbers of subhalos, to infer a probability for the subhalos catalog.

In order to reduce the dimensionality of the problem and enable inference of the collective effects of a large number of low-mass substructures at the statistical level, \cite{Hezaveh_2016b} proposed to use the \gls*{ps} of the lensed deflection field. Subsequently, \cite{DiazRivero_2018a} developed a theoretical general formalism to compute the convergence \gls*{ps} for different subhalo populations from first principles, which was adopted in \cite{DiazRivero_2018b} and \cite{Brennan_2019}. This formalism has been recently expanded to account for \gls*{los} halos in \cite{Sengul_2020}. However, this approach is not directly applicable to observations, because we do not have access to the true displacement field from the data. \citet{Chatterjee_2017}, \citet{Bayer_2018} and \citet{Cyr-Racine_2019} developed statistical formalisms to relate \gls*{ps} of the surface brightness fluctuations in strong lens images to the lens potential fluctuations arising from \gls*{dm} distribution, that contribute to the convergence \gls*{ps}. And \cite{Bayer_2018} applied it to a real observation.

Instead, \cite{Birrer_2017} and \cite{He_2020} employed the residual \gls*{ps} summary statistic, given by the subtraction of a smooth lens model from the data, to constrain the half-mode mass $\mhm$. For the analysis they used approximate Bayesian computation, a likelihood-free inference method based on a rejection algorithm \citep{abc}.
    
Another class of methods that has developed in recent years uses neural networks to measure lens parameters \citep{Hezaveh_2017, PerreaultLevasseur_2017,  Morningstar_2019}, quantifying the structure of gravitational lens potential \citep{Vernardos_2020}, detect individual subhalos \citep{DiazRivero_2020}, and distinguish different types of \gls*{dm} substructure \citep{Alexander_2020}. Still, these methods need lots of data to amortize over all possible variations in lensing systems. In fact, amortized methods learn the posterior for any data, generated by any parameter over the whole range of the prior \citep{Cranmer_2020}. But learning an amortized posterior is unnecessary if only a small range of parameters are consistent with a target observation.

\paragraph*{In this work} we present the first analysis pipeline that combines parametric lensing models with recent neural simulation-based inference developments \citep{Cranmer_2020} to infer the \gls*{dm} mass cutoff scale from a set of realistic simulated galaxy-galaxy strong lenses, by combining their signal. In fact, there are currently around a hundred strong lensing observations suitable for substructure inference, most of which come from the SLACS \citep{SLACS} and BELLS \citep{BELLS} surveys. In the near future, new and future telescopes like JWST \citep{JWST}, ELT \citep{ELT}, Euclid \citep{Euclid_2010, Euclid_2011}, SKA \citep{SKA}, and LSST \citep{LSST} will deliver thousands of very high precision galaxy-galaxy lensing images \citep{McKean_2015}. It is then extremely important to be able to combine the information coming from different observations in the statistical analysis.

For the statistical analysis we employ \gls*{tmnre}. Developed by \citet{Hermans_2020} and \citet{Miller_2020}, \gls*{mnre} is a neural simulation-based inference method that makes it possible to learn the marginal posterior approximation for a specified subset of model parameters of interest directly from the full input data that, in our case, corresponds to the observed lensed images, without the need for hand-crafted summary statistics. This method improves the simulator efficiency and the quality of inference. Moreover, \gls*{mnre} is amortized, which  enables important statistical consistency tests, which would have been extremely expensive with likelihood-based inference, like the expected coverage test we employ in \autoref{subsec:test}. Up to now, this approach has been applied in simplified modeling frameworks: \citet{Hermans_2020} focuses on recovering the Einstein radius of a gravitational lens marginalizing over 15 source and lens mass distribution parameters, whereas \citet{Brehmer_2019} estimates the slope and normalization of a \gls*{cdm} subhalo mass function. Simulation-based inference using neural posterior density estimator and hierarchical inference have also been employed in \citet{Wagner-Carena_2022} to infer the \gls*{cdm} subhalo mass function normalization from a set of strong-lensing images, generated using real galaxy images as a source model, including realistic observational noise effects from \gls*{hst} and accounting for the mean expected convergence from \gls*{los} halos. 

\Gls*{tmnre} is able to \emph{target} the inference to a specific observation at hand rather than amortize over all possible parameters combinations, by successively focusing simulations on the parameter regions that are most relevant for the inference problem \citep{Miller_2021}.
This targeted approach is more efficient when most posterior density is concentrated compared to the prior density, which is the case for lens and source parameters.
This truncation method applied to strong-lensing images was proposed in \citet{Karchev_2021} and used in \citet{Coogan_2020, Coogan_2022} to learn marginal posterior approximations for individual subhalo parameters, marginalizing over lens and source uncertainties given an observation. It has also been recently applied to analysis of the CMB \citep{Cole_2021}.

The main goal of this work is to demonstrate that our \gls*{tmnre} approach is sensitive to the \gls*{hmf} half-mode mass $\mhm$ given a set of HST resolution observations, and it is able to efficiently and accurately infer its statistic.

The paper is structured as follows. In \autoref{model} we describe how we model strong-lens observations with analytic source, lens, and substructure population that accounts for both subhalos and \gls*{los} halos. In \autoref{analysis} we discuss the inference methodology employed in the statistical analysis: \gls*{tmnre}. Finally, we show our results in \autoref{result} and conclude in \autoref{conclusions}. This work paves the way for combining the presented statistical analysis with more realistic strong lensing source models and for future applications to real high-resolution data in upcoming works.

\section{Strong-lensing model}
\label{model}

In this section, we review how we model strong lensing images. In strong-lensing systems the mass distribution of a foreground galaxy gravitationally lenses the light rays coming from a background source, resulting in an arc-like image in the case of an extended galaxy source. Under the assumptions of the thin-lens formalism \citep{Meneghetti_2016}, the lens-plane and source-plane coordinates of a light ray, respectively $\vec{\xi}$ and $\vec{x},$ are related by the simple ray-tracing equation:
\begin{equation}
    \vec{x} = \vec{\xi}-\vec{\alpha}(\vec{\xi}).
\end{equation}
The displacement field $\vec{\alpha}$ can be computed as
\begin{equation}
    \vec{\alpha}(\vec{\xi}) = \frac{4 G}{c^2} \frac{D_{ls}}{D_l D_s} \int \dd[2]{(D_l \vec{\xi}')} \frac{\vec{\xi} - \vec{\xi}'}{|\vec{\xi} - \vec{\xi}'|^2} \Sigma(\vec{\xi}'),
\end{equation}
where we have introduced the angular diameter distance from the observer to the lens $D_l$, from the observer to the source $D_s$, and from
the lens to the source $D_{ls}$. The projected mass density is given by the integral of the 3D lensing mass density $\rho$:
\begin{equation}
    \Sigma(\vec{\xi}) = \int \dd{z} \rho(\vec{\xi}, z),
\end{equation}
where $z$ is the coordinate perpendicular to the lens plane. It is also useful to define the convergence $\kappa$ in terms of the critical surface density $\Sigma_{\mathrm{cr}, l}$ on the lens plane as:
\begin{equation}
    \kappa(\vec{\xi})=\cfrac{\Sigma(\vec{\xi})}{\Sigma_{\mathrm{cr}, l}},
    \quad \Sigma_{\mathrm{cr}, l}\equiv\cfrac{c^2}{4\pi G}\cfrac{D_s}{D_l D_{ls}},
\end{equation}
where $c$ is the speed of light and $G$ is the gravitational constant. It can be shown that the convergence $\kappa$ is related to the trace of the Jacobian of the lensing transformation, and it represents the lens mass distribution. 

Strong-lensing systems are then represented by two main ingredients: the lens model, which describes the total mass distribution of the lens, and the source model, which describes the surface brightness profile of the background source. It is common to split the lens model into a macroscopic smooth component (main lens and external shear) and a substructure\footnote{Throughout our work, we use the terms `small-scale structures', `substructures', and `low-mass halos' when considering both subhalos of the main lens and \gls*{los} halos.} component, due to subhalos and \gls*{los} halos. Each ingredient can be directly superimposed by summing their respective displacement fields in the lens plane:
\begin{equation}
    \vec{\alpha} = \vec{\alpha}_\mathrm{lens} + \vec{\alpha}_\mathrm{ext} + \sum_{i=1}^{N_{\mathrm{sub}}} \vec{\alpha}_\mathrm{sub,i} + \sum_{i=1}^{N_{\mathrm{los}}} \vec{\alpha}_\mathrm{los,i}.
\end{equation}

In the following subsections, we will describe each component of the model, which we summarize in \autoref{tab:model}.

\begin{table}
\caption{Summary of model parameters used for the simulated images in this work. When a prior distribution is not specified, the parameter is fixed to the true value.}
\label{tab:model}
\begin{center}
\resizebox{\columnwidth}{!}{%
    \begin{tabular}{l l l c }
        \toprule
        Parameter & {True value} & Prior & Description \\
        \midrule
        \textbf{Main lens} &  &  &  SPLE \\
        $r_\mathrm{Ein}$ [\si{\arcsecond}] &  & $\Uniform(1.,2.)$  & Einstein radius \\
        $\xi_{0,x}$ [\si{\arcsecond}] &  & $\Uniform(-0.2,0.2)$  & lens center x axis \\
        $\xi_{0,y}$ [\si{\arcsecond}] &  & $\Uniform(-0.2,0.2)$  & lens center y axis \\
        $q_l$ &  & $\Uniform(0.1,1.)$ & axis ratio \\
        $\phi_l$ [\si{\radian}]&  & $\Uniform(0,2\pi)$  & rotation angle \\
        $\gamma$ & 2.1 & -  & slope \\
        $z_\mathrm{lens}$ & 0.5 & -  & lens redshift \\
        \midrule
        \textbf{External shear}  &  &  &   \\
        $\gamma_1$ &  & $\Uniform(-0.05,0.05)$  & $1^{\mathrm{st}}$ component\\
        $\gamma_2$ &  & $\Uniform(-0.05,0.05)$   & $2^{\mathrm{nd}}$ component\\
        \midrule
        \textbf{Source} &  &  &  Sérsic \\
        $I_e$ &  & $\Uniform(0.,4.)$  & surface intensity \\
        $r_e$ [\si{\arcsecond}] &  &  $\Uniform(0.1,2.5)$ & effective radius \\
        $x_0$ [\si{\arcsecond}] &  & $\Uniform(-0.1,0.1)$ & source center x axis\\
        $y_0$ [\si{\arcsecond}] &  & $\Uniform(-0.1,0.1)$ & source center y axis\\
        $q_s$ &  & $\Uniform(0.1,1.)$ & axis ratio\\
        $\phi_s$ [\si{\radian}] &  &  $\Uniform(0,2\pi)$  & position angle\\
        $n$&  &  $\Uniform(0.1,4.)$  & index\\
        $z_\mathrm{src}$ & 2 & -  & source redshift\\
        \midrule
        \textbf{Subhalos} & & &  tNFW \\
        $\vec{p}$ [\si{\arcsecond}] & {$\in[-2.5, 2.5]$} & $\Uniform_{2\mathrm{D}}(-2.5, 2.5)$ & position\\
        $m_{200}$ [\si{\solmass}] & {$\in[10^7, 10^{10}]$} & \cite{Giocoli_2010} & virial mass\\
        $c_{200}$ & 15. & - & concentration \\
        $\tau$ & 6. & - & truncation\\
        \midrule
        \textbf{\gls*{los} halos} &  &  &  projected tNFW \\
        $\vec{p}$ [\si{\arcsecond}] & {$\in[-2.5, 2.5]$} & $\Uniform_{2\mathrm{D}}(-2.5, 2.5)$ & position\\
        $m_{200}$ [\si{\solmass}] & {$\in[10^7, 10^{10}]$} & \cite{Tinker_2008} & virial mass\\
        $z_{\mathrm{LOS}}$ & {$\in[0, z_\mathrm{src}]$} &  \cite{Tinker_2008} & LOS redshift\\
        $c_{200}$ & 15. & - & concentration \\
        $\tau$ & 6. & - & truncation\\
        \midrule
        \textbf{\gls*{wdm}} &  & &  \\
        $\mhm$ [\si{\solmass}] & & $\log\Uniform(10^7, 10^{10})$ & half-mode mass\\
        \bottomrule
    \end{tabular}
}
\end{center}
\end{table}

\subsection{Main lens model}
We model the lens mass distribution smooth component with a \gls*{sple} lens \citep{Suyu_2009} plus external shear. The latter accounts for matter in the lens surroundings and describes large-scale effects constant across the image. For a detailed description of these two ingredients, we refer the reader to \cite{Karchev_2021}. We end up with eight parameters in total that we collect in the vector $\btheta_l \equiv \{r_\mathrm{Ein}, \xi_{0,x}, \xi_{0,y}, q_l, \phi_l, \gamma, \gamma_1, \gamma_2\}$, the first six from the \gls*{sple} for the main-lens mass distribution and the last two for external shear. 

When simulating data (see \autoref{subsec:data}), we always use the same \gls*{sple} slope that produced the mock observation\footnote{
    Throughout our work,  we use the terms “simulated data” for data used during inference and “mock observations” for the simulated data that we analyse.
    } 
we are analysing (as fixed in \autoref{tab:model}) instead of inferring it for simplicity.\footnote{
    In principle, inferring the slope is possible, but it requires more training data and leads to increased uncertainties in both lens and source parameters.
    }

\subsection{Source model}
\label{subsec:source}
To model the surface brightness of the source galaxy, we adopt a Sérsic profile \citep{Sersic_1963}. The surface brightness distribution is given by 
\begin{equation}\label{eq:sersic}
    \beta(x, y)=I_e \exp{-k_n\left[\left(\cfrac{r(x, y)}{r_e}\right)^{1/n}-1\right]},
\end{equation}
where $I_e$ is the surface intensity at the half-light radius $r_e$, $r(x, y)=\sqrt{r_x^2+r_y^2}$ is the elliptical radial coordinate, and the normalization $k_n$ depends on the index $n$. We give more details about the Sérsic profile modeling and parameters in appendix \ref{apx:sersic}. In total, the analytic source is parametrised with seven variables that we collect in the vector $\btheta_s\equiv\{I_e, r_e, x_0, y_0, q_s, \phi_s, n\}$.

\subsection{Small-scale structures model}
\label{subsec:substructure}

\begin{figure}
    \centering
    \includegraphics[width=\linewidth]{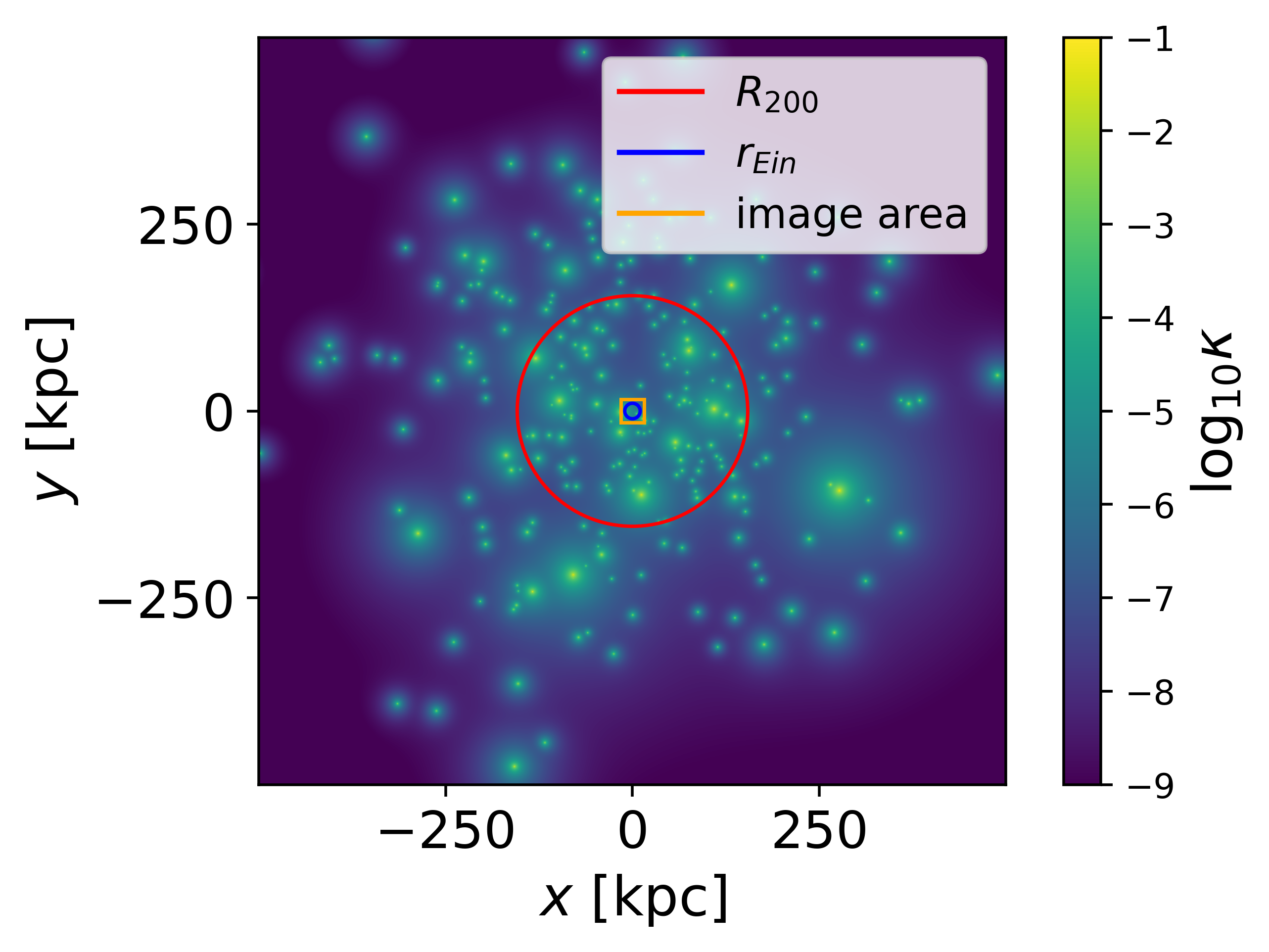}
    \caption{
        Convergence map for a \gls*{cdm} subhalo population in the adopted mass range. The convergence map shows how the deflecting mass from all the subhalo lenses is distributed. The full map size is $1 \times 1\ \si{\Mpc}$. We mark in red the virial radius of the main lens halo, in blue its Einstein radius, and in orange the $5 \times 5\ \si{\arcsec}$ lensing image area.
    }
    \label{fig:sub-convergence}
\end{figure}

\begin{figure}
    \centering
    \includegraphics[width=\linewidth]{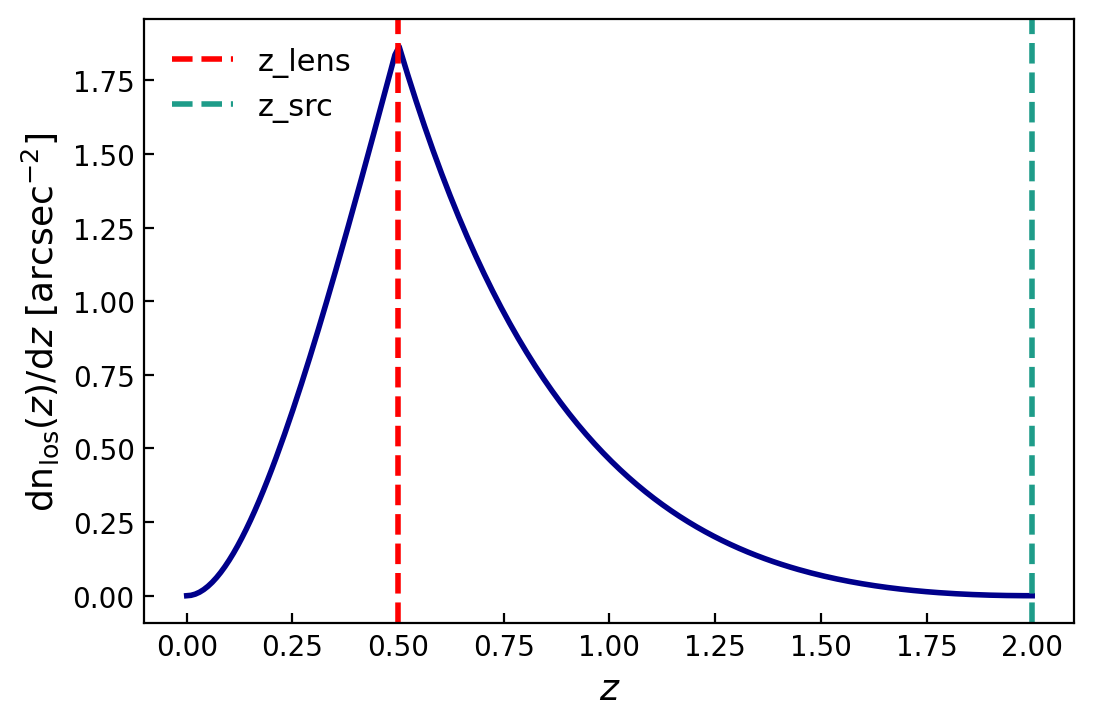}
    \caption{\gls*{los} halos distribution in redshift for our source and lens redshifts configuration, described in \autoref{tab:model}.}
    \label{fig:los-z-dist}
\end{figure}

Substructures can be divided into two categories: subhalos that orbit around the main halo at the lens redshift, and \gls*{los} halos distributed between the source and the observer. \Gls*{los} halos are a more direct probe of free-streaming-induced small-scale structure suppression, because they are less affected by baryonic processes and environmental effects, such tidal stripping interactions with the main halo \citep{Despali_2018}. For this reason and the fact that they are expected to be more abundant than subhalos in a lensing system \citep{Despali_2018, He_2021}, it is very important to model them as well, in order to correctly estimate the collective effects of all substructures on the lensing image. 

\subsubsection{Density profile}
To model the density profiles of small-scale \gls*{dm} halos we adopt the smoothly truncated universal 3D mass density profile from \cite{Baltz_2009}:
\begin{equation}
    \rho_{\mathrm{tNFW}}(r)= \cfrac{\rho_s}{r/r_s (1+r/r_s)^2}  \cfrac{1}{1+(r/r_t)^2}.
\end{equation}
Here $r$ is the three-dimensional distance from the center of the halo, $\rho_s$ and $r_s$ are respectively the scale density and scale radius that specify an \gls*{nfw} profile \citep{nfw}, and $r_t\equiv\tau r_s$ is the tidal truncation radius that depends on the history of the subhalo. Typical values of the truncation scale $\tau$ range from $4-10$ for spherically symmetric lenses \citep{Gilman_2019b, Cyr-Racine_2019}; we fix $\tau=6$ for simplicity. 
Compared to the standard \gls*{nfw} form, which has an infinite total mass, the \gls*{tnfw} contains an additional truncation term that makes the profile decay as $r^{-5}$ for large radii, resulting in a finite total mass given by:
\begin{equation}
    m_\tau=4\pi \rho_s r_s^3 \frac{\tau^2}{(\tau^2+1)^2}[(\tau^2-1)\ln\tau+\tau\pi-(\tau^2 +1)].
\end{equation}
With a fixed truncation scale, the \gls*{tnfw} profile is fully determined by the same parameters that determine the \gls*{nfw} profile: the virial mass $m_{200}$\footnote{We parameterize subhalos by what would be their mass up to the virial radius $r_{200}$ using the untruncated profile, with the same central density $\rho_s$ and scale radius $r_s$ as the truncated one.} and the concentration $c_{200}=r_{200}/r_s$ of the halo. The latter measures how concentrated the mass of a halo is and fixes the density normalization; in principle, it varies from one subhalo to the next and shows dependencies on mass and redshift of the main halo. In this paper, instead of adopting a concentration-mass relation, we fix $c_{200}=15$ in accordance with \cite{Richings_2020}. We would like to note that accounting for the scatter in the mass-concentration relation might boost the expected lensing signal from a low-mass halo \citep{Amorisco_2021}.

\gls*{los} halos are also modeled with a \gls*{tnfw} profile following the prescription by \cite{Sengul_2020}, which shows how to treat halos along the line-of-sight as effective subhalos on the main-lens plane, with a modified scale radius and mass. We give more details on this procedure in appendix \ref{apx:los}. For \gls*{los} halos we adopt the same concentration and truncation scale values used for subhalos.

The equations for calculating the displacement field of a \gls*{tnfw} halo, given its mass and position, are fully elaborated by \citet[Appendix A]{Baltz_2009}.

\subsubsection{Mass and spatial distributions}
\label{subsubsec:sub-dist}

We sample subhalo masses from the \gls*{cdm} mass function of \cite{Giocoli_2010}:
\begin{equation}\label{eq:shmf}
    \frac{1}{M}\frac{\dd n_\mathrm{sub}( m_{200}, z)}{\dd\log m_{200}} \propto (1+z)^{1/2}m_{200}^{\alpha} \exp\left[-\beta\left(\frac{m_{200}}{M}\right)^3 \right],
\end{equation}
where $M$ is the main halo's mass and $m_{200}$ the subhalo mass \footnote{
    The total mass of the lens galaxy is described by the Einstein radius of the system, a very well-constrained parameter in lensing inference analyses. For the purpose of describing subhalos, we need to be able to map the measured properties of the lens (the Einstein radius $r_\mathrm{Ein}$) onto the properties of the host halo (the mass $M$). For simplicity, we compute the mass of the host halo transforming the Einstein radius distance measure into a mass measure. We would like to point out a similar approach from \citet{Brehmer_2019}, where they relate the central velocity dispersion of a singular isothermal ellipsoid lens mass distribution profile to the virial mass of the host halo.
}. We use the normalization, slope and exponential cutoff of the subhalo mass function from \citet{Despali_2017}. The expected number of subhalos in a given mass interval for the lens halo system can be computed by integrating the mass function over that interval.

For \gls*{los} halos masses we use the \cite{Tinker_2008} \gls*{cdm} halo mass function assuming an overdensity with respect to the critical density of the Universe at the epoch of analysis of $\Delta=200$. For both subhalos and \gls*{los} halos we adopt the following mass range, with $m_{200, \mathrm{min}}= 10^7\ \si{\solmass}$ and $m_{200, \mathrm{max}}= 10^{10}\ \si{\solmass}$. The upper limit is chosen based on the assumption that more massive halos would be visible and could therefore be modeled independently. The lower one is fiducial, and we plan on investigating more the sensitivity of our inference in the future.

The spatial distribution of subhalos has been shown to follow an Einasto profile \citep{Springel_2008}. However, since the virial radius of a typical main lens halo is much larger than its Einstein radius, and hence, than the image plane, we approximate the distribution to be uniform in the lensing image area. Still, we derive the total number of expected subhalos within the image via the Einasto fit of \citet{Despali_2017}. We find that on average $\bar{n}_\mathrm{sub}=4$ subhalos fall within the lensing image area in our adopted lensing configuration and mass range. When generating a simulated image, we draw the number of subhalos from $\operatorname{Poisson}(\bar{n}_\mathrm{sub})$, we then sample their masses from the subhalo mass function in \autoref{eq:shmf} and sample their projected positions uniformly over the lensing image area.
In \autoref{fig:sub-convergence} we show the convergence map for one realization of our subhalo population.

\gls*{los} halos are rendered in a double-cone geometry with the lensing image area as an opening angle, and closing angle such that the cone closes at the source redshift, as described in \citet[Figure 3]{Sengul_2020}.
We infer the number of detectable \gls*{los} halos by integrating their mass function in the mass range adopted for the analysis and within the double-cone volume
\begin{equation}
    \bar{n}_{\mathrm{los}}=\int_0^{z_\mathrm{src}}\int_{m_{200, \mathrm{min}}}^{m_{200, \mathrm{max}}} n_\mathrm{los}(m_{200}, z)\dd{m_{200}} \frac{\dd{V}}{\dd{z}}\dd z.
\end{equation}
On average we get $\bar{n}_{\mathrm{los}}=260$ \gls*{los} halos projected in our lens plane. Similarly to what we do with the subhalo population, when generating simulated images, we draw the number of LOS halos from $\operatorname{Poisson}(\bar{n}_\mathrm{los})$, we then sample their masses and redshift from the \cite{Tinker_2008} halo mass function and sample their projected positions uniformly over the lensing image area.
In \autoref{fig:los-z-dist} we show the distribution of \gls*{los} halos in redshift for our lens and source redshifts configuration.

Finally, we label the vector of all substructure parameters with $\btheta_h \equiv \{\msub, \psub,  \mlos,  \plos, \zlos\}$, where we use bold letters to denote arrays (e.g.\ $\msub$ is an ordered set of masses, one for each simulated subhalo) and bold letters with an arrow to indicate arrays of vectors (e.g.\ $\psub$ is an ordered set of positions in the lens plane, one for each simulated subhalo).

\subsection{Modelling free-streaming effects in WDM}
\label{subsec:free-streaming}

\begin{figure*}
    \centering
    \includegraphics[width=\linewidth]{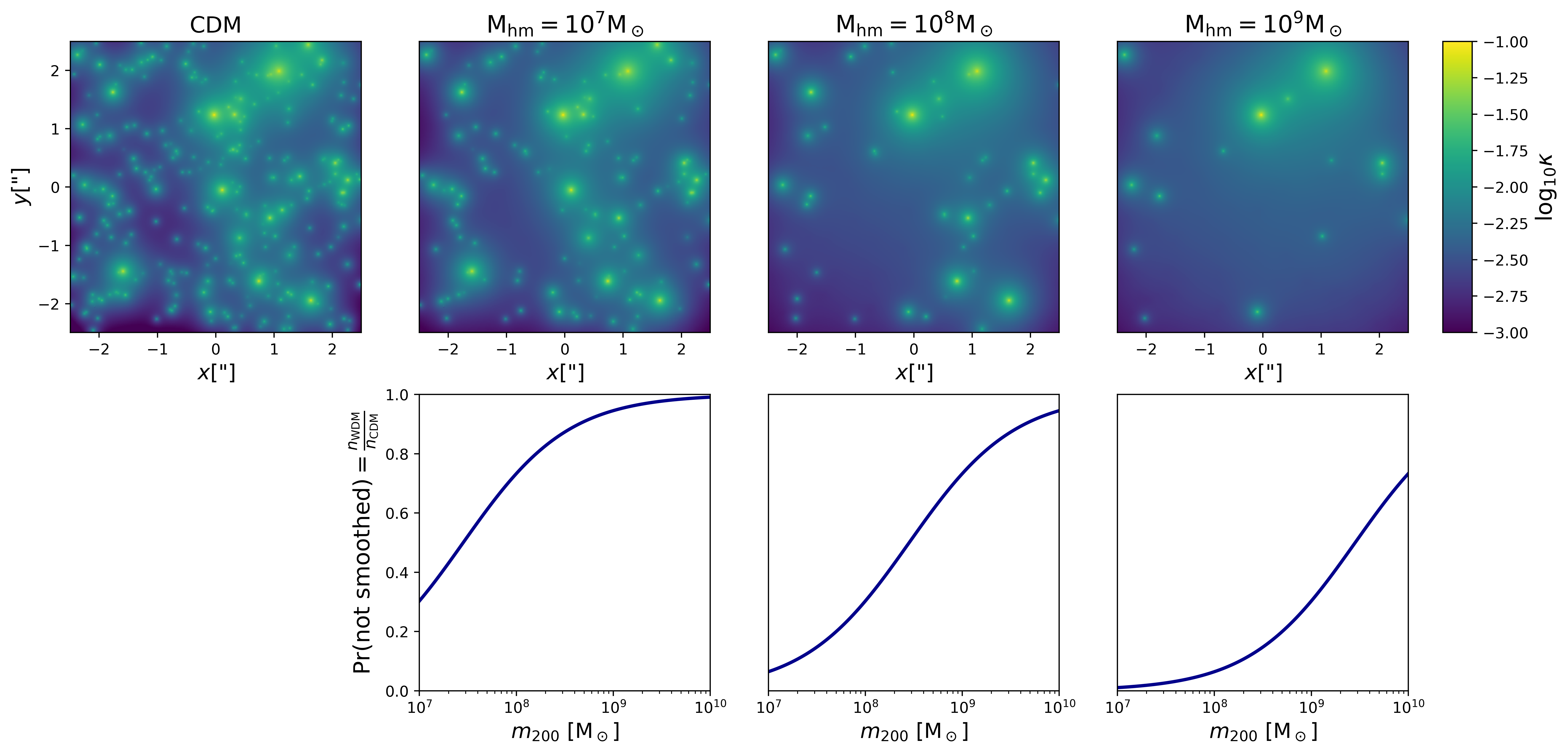}
    \caption{
    \textit{Top}: Convergence maps for a population of \gls*{los} halos with masses sampled from the \gls*{cdm} \gls*{hmf} in the adopted mass range (\autoref{tab:model}) and projected in the lens plane following the prescription from \citet{Sengul_2020}.
    In the second to fourth columns, we imitate the effect of WDM with different cutoff masses (as labeled in the titles) via our smoothing scheme. \textit{Bottom}: We show the probability with which \gls*{los} halos do not get smoothed, equal to the ratio between the WDM and the CDM HMF (\autoref{eq:cutoff}). The smoothing is stochastic, so for each realization of the smoothing different halos are smoothed.
    }
    \label{fig:smoothing}
\end{figure*}

The free-streaming effects of \gls*{wdm} are well described in terms of the half-mode wavelength $\lambda_{\mathrm{hm}}$, which corresponds to the scale at which the \gls*{dm} transfer function falls to half the \gls*{cdm} transfer function. We can define the half-mode mass as the mass contained within a radius of the half-mode wavelength:
\begin{equation}
\mhm=\frac{4\pi\Omega_{\mathrm{M}}\rho_{\mathrm{crit}}}{3}\left(\frac{\lambda_{\mathrm{hm}}}{2}\right)^3,
\end{equation}
where $\Omega_{\mathrm{M}}$ is the matter density parameter and $\rho_{\mathrm{crit}}$ is the critical density of the Universe.
Following \cite{Schneider_2012}, the half-mode wavelength,
\begin{equation}
    \lambda_{\mathrm{hm}}=2\pi\alpha_{\mathrm{hm}}\left(2^{\nu/5}-1\right)^{-1/(2\nu)},
\end{equation}    
is the scale below which the initial density perturbations are completely erased, with $\nu= 1.12$ and, assuming that all \gls*{dm} is warm,
\begin{equation}
\alpha_{\mathrm{hm}}=0.049\left(\frac{m_\mathrm{WDM}}{\si{\keV}}\right)^{-1.11}\left(\frac{\Omega_\mathrm{DM}}{0.025}\right)^{0.11}\left(\frac{h}{0.7}\right)^{1.22}h^{-1}\si{\Mpc}.
\end{equation}

We then have a one-to-one mapping between the mass of the \gls*{wdm} particle and the half-mode mass.
For strong lensing, the half-mode mass can be thought of as an effective cutoff mass below which the \gls*{dm} mass function is strongly suppressed. To model this suppression in the \gls*{wdm} mass function we adopt for both subhalos and \gls*{los} halos the functional form from \cite{Lovell_2020}:
\begin{equation}\label{eq:cutoff}
    \frac{n_{\mathrm{WDM}}}{n_{\mathrm{CDM}}}=\left(1+\left(\alpha\frac{\mhm}{m_{200}}\right)^\beta\right)^\gamma,
\end{equation}
with best fit parameters $\alpha=4.2$, $\beta=2.5$, and $\gamma=0.2$ for subhalos, $\alpha=2.3$, $\beta=0.8$, and $\gamma=1$ for central halos.

\subsubsection{Smoothing substructures}
\label{subsubsec:smoothing}

The observational signature of \gls*{wdm} is, thus, the absence of small-scale structures. However, in the current parameterization, this is accompanied by the removal of the corresponding mass enclosed in them, whereas in reality the mass will still be present but will be diffused throughout the smooth main halo. This effect is manifested in a correlation between the half-mode mass and the main-halo Einstein radius: suppressing more substructure leads to an increase in the inferred Einstein radius since the total mass of the system (within the image) is tightly constrained by the size of the observed ring (or arcs).

We introduce a prescription for dealing with this degeneracy, which well captures the physical reality of structure suppression due to free streaming. 
Halos that should be suppressed are not present because the \gls*{dm} particles that should make them up are freely streaming, and their mass is therefore more diluted throughout the main halo.
Therefore, rather than removing or adding substructures as a response to a changing cutoff, we still sample substructures from the \gls*{cdm} mass function, but we smooth the displacement field generated by halos that should be suppressed based on the aforementioned prescription by \cite{Lovell_2020} to hide their lensing signature. In other words, each sampled small-scale halo has a probability equal to the ratio between the \gls*{wdm} and the \gls*{cdm} \gls*{hmf} (\autoref{eq:cutoff}) of not being smoothed. 

We then effect the smoothing by convolving the deflection field of each individual sub-/\gls*{los} halo with a radially-symmetric filter
\begin{equation}
f \propto 1 - \exp\left(-\left(\frac{r}{r_{\mathrm{smooth}}}\right)^{n_{\mathrm{smooth}}}\right).
\end{equation}
This filtering preserves the far-field lensing signature of the halo, which is only determined by its total mass.
By default, we choose the smoothing scale to be equal to the halo virial radius: $r_\mathrm{smooth}=r_{200}$, and the smoothing exponent $n_\mathrm{smooth}=2$.

In the top row of \autoref{fig:smoothing} we visualize the convergence maps in the lens plane for the same realization of \gls*{los} halos drawn from \gls*{cdm} distributions (panel 1), and with different cutoff masses implemented with our smoothing scheme (panels 2-4). In the bottom row, we show how we decide to smooth the lensing signature of certain halos based on the ratio between the \gls*{wdm} and the \gls*{cdm} \gls*{hmf} (\autoref{eq:cutoff}). 

\section{Statistical analysis}
\label{analysis}

Constraining the fundamental properties of \gls*{dm} by characterizing the population of \gls*{dm} halos in a strong lensing image is an extremely difficult problem since the signal we are interested in has a sub-percent level influence on images dominated by statistical noise. The problem is further complicated by the large differences between images of different lensing systems.

Our ultimate goal is to compute the marginal posterior $p(\vartheta|\bx)$ for a single parameter of interest $\vartheta=\mhm$, the half mode mass, given an observation $\bx$, for which we have the generative model
\begin{equation}
\label{eq:model}
    p(\bx, \btheta_s, \btheta_l, \btheta_h, \vartheta)
    = p(\bx|\btheta_s, \btheta_l, \btheta_h, \vartheta) p(\btheta_s)p(\btheta_h|\btheta_l, \vartheta) p(\btheta_l) p(\vartheta).
\end{equation}
The first factor on the right-hand side is the simulator, while the other factors denote the priors on the various source, lens, and \gls*{dm} substructures parameters as listed in \autoref{tab:model}. 

Therefore, in order to derive $p(\vartheta|\bx)$ we need to marginalise over all the nuisance parameters $\boldeta\equiv\{\btheta_l, \btheta_s, \btheta_h\}$:
\begin{equation}\label{eq:post}
     p(\vartheta|\bx) = 
     \frac{p(\bx|\vartheta)}{p(\bx)} p(\vartheta) =
     \frac{\int \dd{\boldeta} p(\boldeta) p(\bx|\vartheta, \boldeta)}{p(\bx)} p(\vartheta).
\end{equation}
This is a very high-dimensional and multi-modal integral, even for simple analytical lens and source models, due to the large population of interchangeable substructures. Therefore, it is intractable, which renders likelihood-based inference infeasible in this case.

Instead, we approximate $p(\vartheta|\bx)$ using simulation-based inference with amortized approximate ratio estimators \citep{Hermans_2020}. In particular, we employ the \gls*{tmnre} algorithm developed by \citet{Miller_2020, Miller_2021} and implemented in the package \swyft\footnote{\url{https://github.com/undark-lab/swyft}.}.

\subsection{Marginal neural ratio estimation}
\label{subsec:mnre}

\gls*{mnre} \citep{Miller_2020} sets up a classification problem which produces an estimate $\hat{r}(\bx,\vartheta)$ of the marginal likelihood-to-evidence ratio:
\begin{equation}\label{eq:ratio}
    r(\bx,\vartheta)  \equiv \frac{p(\vartheta|\bx)}{p(\vartheta)} = \frac{p(\bx|\vartheta)}{p(\bx)} = \frac{p(\bx, \vartheta)}{p(\bx)p(\vartheta)}.
\end{equation}
Given the prior $p(\vartheta)$, the ratio estimator $\hat{r}(\bx,\vartheta)$ may then be used as a surrogate model to draw samples from an approximate posterior $\hat{p}(\vartheta|\bx) = \hat{r}(\bx,\vartheta) p(\vartheta)$.
In order to estimate the ratio, the strategy is to train a neural network $d_\phi(\bx,\vartheta)$, where $\phi$ are the network weights, via stochastic gradient descent. The network is parameterised as a binary classifier to discriminate between two hypotheses labeled by the binary variable $C$. In the first one, with class label $C=1$, the observation $\bx$ and the parameter of interest $\vartheta$ are drawn jointly from the parameter prior and model: $\bx, \vartheta \sim p(\bx, \vartheta)$. In the second one,  with class label $C=0$, they are sampled marginally: $\bx, \vartheta \sim p(\bx)p(\vartheta)$. We sample from the two classes with equal probability, enforcing the outcome of the binary variable $C$ to be random. 
To train the network we use the binary-cross entropy loss function:

\begin{equation}\label{eq:bce}
\begin{split}
    \ell[d_\phi(\bx,\vartheta)] &= -\int \dd{\bx} \dd{\vartheta} \left\{ p(\bx, \vartheta) \log d_\phi(\bx,\vartheta) \right. \\
    &\hspace{1.5cm} \left. + p(\bx) p(\vartheta) \log\left[ 1 - d_\phi(\bx,\vartheta) \right] \right\} .
\end{split}
\end{equation}

The loss functional defined in \autoref{eq:bce} is minimized when the output of the neural network $d_\phi(\bx,\vartheta)$ corresponds to the probability of the class with label $C=1$:
\begin{equation}
    d_\phi(\bx,\vartheta) = p(C = 1 | \bx, \vartheta) = \frac{p(\bx, \vartheta)}{p(\bx, \vartheta) + p(\bx) p(\vartheta)} \equiv \sigma[ \log \hat{r}(\bx, \vartheta)] \, ,
\end{equation}
so we can express the ratio estimator $\hat{r}(\bx,\vartheta)$ in terms of the binary classifier $d_\phi(\bx,\vartheta)$ using the sigmoid function $\sigma(y) \equiv 1 / (1 + e^{-y})$.

Marginalization over nuisance variables $\boldeta$ is done implicitly since the data will incorporate the variance from the nuisance parameters, but the inference procedure estimates only the marginal likelihood-to-evidence ratio. In other words, parameters to be marginalized over are sampled during training data generation, but not shown to the binary classifier $d_\phi(\bx,\vartheta)$. As a result, the trained network effectively learns an estimate of the marginal likelihood-to-evidence ratios $\hat{r}(\bx,\vartheta)$, which we can use to evaluate the marginal posterior for the parameter of interest directly (if the prior PDF is known) or obtain samples otherwise.

\subsection{Truncated marginal neural ratio estimation}
\label{subsec:tmre}

Formally, inferring the marginal posterior for the substructure population parameter of interest would require marginalizing over all the source, lens, and substructure realizations compatible with all possible strong lensing images. However, sampling lens and source parameters from their priors would require a very large amount of training data and a more complex network architecture when using neural ratio estimation. This has been attempted only in \citet{Brehmer_2019} to infer the slope and normalization of the \gls*{hmf}. In order to reduce the complexity of the problem and fully exploit available information in the data with limited computational resources, we propose to target one image at a time, focusing simulations and the network training on a specific observation of interest. Thanks to \swyft, we implement this with a truncation scheme.

\gls*{tmnre} generates a sequence of likelihood-to-evidence ratio estimators on both nuisance and parameters of interest for a specific observation $\bx$. In multiple inference rounds, the proposal distribution for nuisance parameters is updated and constrained, based on these ratio estimators, in order for the training data to match each round more closely the observation of interest $\bx$ (this can be visually appreciated in \autoref{fig:targeted_data}, which will be discussed in more details in \autoref{subsec:constrain}). 

The procedure for the truncation scheme is the following. In the first inference round, we generate training data sampling the nuisance parameters from the initial prior $p(\boldeta)$. Then, in each round, we constrain the proposal distribution $p_\Gamma(\boldeta)$ for the parameters we want to marginalize over to a region $\Gamma$ where the nuisance parameters are more likely to have generated $\bx$ based on the ratio estimator trained in that round. In particular, we estimate the new region $\Gamma$ by very conservatively truncating the previous proposal distribution $p_\Gamma(\boldeta)$ in the region where the ratio estimator exceeds a predetermined threshold. We set the threshold hyperparameter to $\epsilon = 10^{-5}$, which, in case of a Gaussian posterior, corresponds to truncating at $\sim 4.78\ \sigma$ \citep{Miller_2021}. We obtain the final proposal distribution for our nuisance parameters when the region $\Gamma$ does not change significantly anymore between rounds.

\medskip
In this work we target with \gls*{tmnre} a restricted set of the nuisance parameters $\boldeta$: namely, those of the analytic smooth lens and source models, $\btheta_l$ and $\btheta_s$, while leaving halo parameters, $\btheta_h$ unconstrained.

\medskip
To summarize, thanks to \gls*{tmnre}, the overall analysis strategy splits into the following steps:
\begin{enumerate}
    \item[{1.}] Train an inference network on an image $\bx$ to constrain the source and lens parameters, $\btheta_s$ and $\btheta_l$, within ranges consistent with the observation. We then generate targeted training data based on this constrained model.
    \item[{2.}] Train an inference network to learn the marginal likelihood-to-evidence ratio for our parameter of interest, the half mode mass $\mhm$, on the targeted training data.
\end{enumerate}

Similarly to the reasoning behind the approximate Bayesian computation rejection algorithm, which discards sampled parameters values if the generated data is too different from the observed data, we justify this approach by noting that parameters that do not produce observations similar to $\bx$ will not contribute to the integral in \autoref{eq:post}. Restricting the input parameters in this way immensely reduces the variability of simulated data, which allows us to use simpler network architectures and fewer training examples in the next step. As a result, the inference is now \emph{targeted} to the specific observation at hand rather than amortized over all the possible lens/source combinations from the full prior. We would like to point out that the inference is still locally amortized in the constrained proposal distribution region, and this enables empirical test of the inference result (see \autoref{subsec:test}).

\section{Results}
\label{result}

In this section, we show our results. First, we describe the simulated data in \autoref{subsec:data} and the inference network architectures in \autoref{subsec:nn}. We then show how we constrain the lens and source parameters in \autoref{subsec:constrain}. Next, we show our results for the cutoff mass and describe how we can combine the information from different strong lensing images in \autoref{subsec:dm}. In the same subsection, we show our results on the \gls*{dm} mass. Finally, we directly assess the statistical behaviour of the trained neural networks in \autoref{subsec:test}.

\subsection{Mock data generation}
\label{subsec:data}

\begin{figure}
    \centering
    \includegraphics[width=\linewidth]{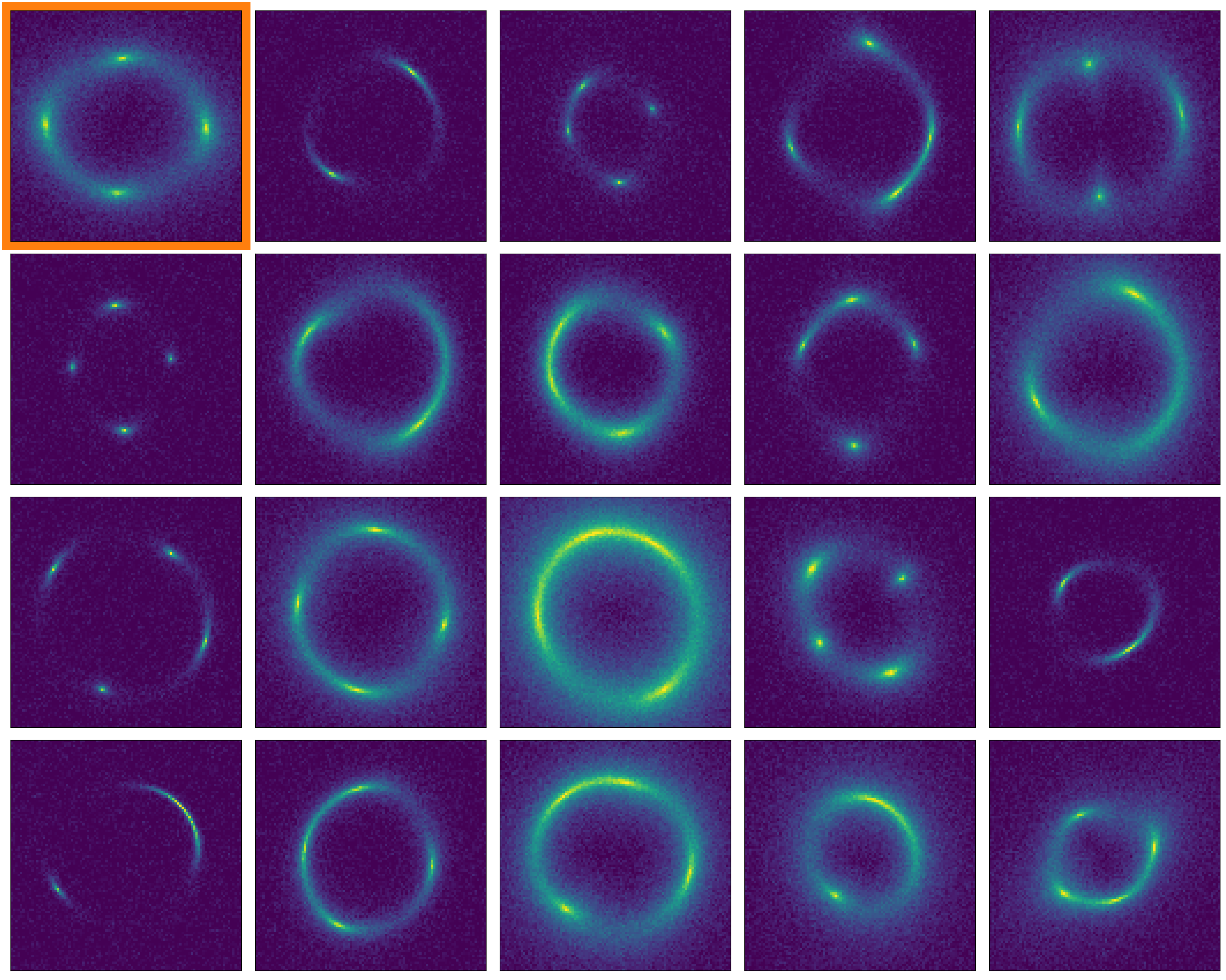}
    \caption{We present a gallery of twenty mock strong-lensing images we use as target observations. These mock observations have been generated with arbitrary lens and source parameters drawn from the initial prior in \autoref{tab:model}. Their peak \gls*{snr} is $\num{\sim 30}$, representative of \gls*{hst} data. We analyse these images by first constraining their lens and source parameters proposal distribution in \autoref{subsec:constrain}. Then, we combine them in order to infer the cutoff mass scale in \autoref{subsec:dm}. For the first one (upper left corner, framed in orange) of these images we show our results of the first part of the pipeline (\autoref{subsec:constrain}) in \autoref{fig:bounds}, \autoref{fig:targeted_data}, and \autoref{fig:lens_source_post}.
}
\label{fig:mock}
\end{figure}

We want our simulations to be representative of \gls*{hst} data, in order to demonstrate that our pipeline is in principle able to extract the signal of interest from them. We adopt a pixel scale of \SI{0.05}{\arcsecond}, being slightly larger than the expected \SI{0.04}{\arcsecond}, which allows us to disregard its \gls*{psf} \citep{HST} for simplicity. The size of the images is $100 \times 100$ pixels, so they cover an area of $5 \times 5\ \si{\arcsecond}$ on the sky. 
Initially, we generate the mock data with a resolution 10-times higher and then downsample it to the adopted resolution by local averaging, effectively simulating integration of the light across the pixel areas.

We model the instrumental effects by simply assuming a Gaussian and uncorrelated pixel noise. The noise level $\sigma$ is set so that the peak \gls*{snr} ratio of the image is $\num{\sim 30}$ (after downsampling), representative of \gls*{hst} data. Then, given a modeled flux, our simulator is given by:
\begin{equation}
    p(\bx|\btheta_s, \btheta_l,\btheta_h, \vartheta)
    = \mathcal{N}(\bx|\mathrm{obs}(\btheta_s, \btheta_l, \btheta_h, \vartheta), \sigma^2).
\end{equation}
We leave to future works to account for the correct modeling of the \gls*{psf} and correlated pixel noise, which are fundamental in order to correctly conduct substructure studies in strong gravitational lensing images.

In \autoref{fig:mock} we show a gallery of twenty mock strong-lensing images we use as target observations. These mock observations have been generated with arbitrary lens and source parameters drawn from the initial prior in \autoref{tab:model}. Their peak \gls*{snr} is $\num{\sim 30}$, representative of \gls*{hst} data.

\subsection{Inference network architecture}
\label{subsec:nn}

The inference neural network used to perform \gls*{tmnre} is split into two different components: an embedding network $C_\phi(\bx)$ and a binary classification network. 
The embedding network compresses data into a low-dimensional feature vector, estimating the best possible summary statistics from the full input image. 
The binary classification network is the marginal classifier that performs the actual ratio estimation. It passes the featurized observational data concatenated with the parameter of interest into a \gls*{mlp} to estimate the marginal likelihood-to-evidence ratios.
The network architecture can be expressed as:
\begin{equation}
    d_\phi(\bx, \vartheta) = \mathrm{MLP}_\phi(\text{features} = C_\phi(\bx), \vartheta) = \sigma[ \log \hat{r}(\bx, \vartheta)].
\end{equation}

\begin{figure}
\centering
\includegraphics[width=\linewidth]{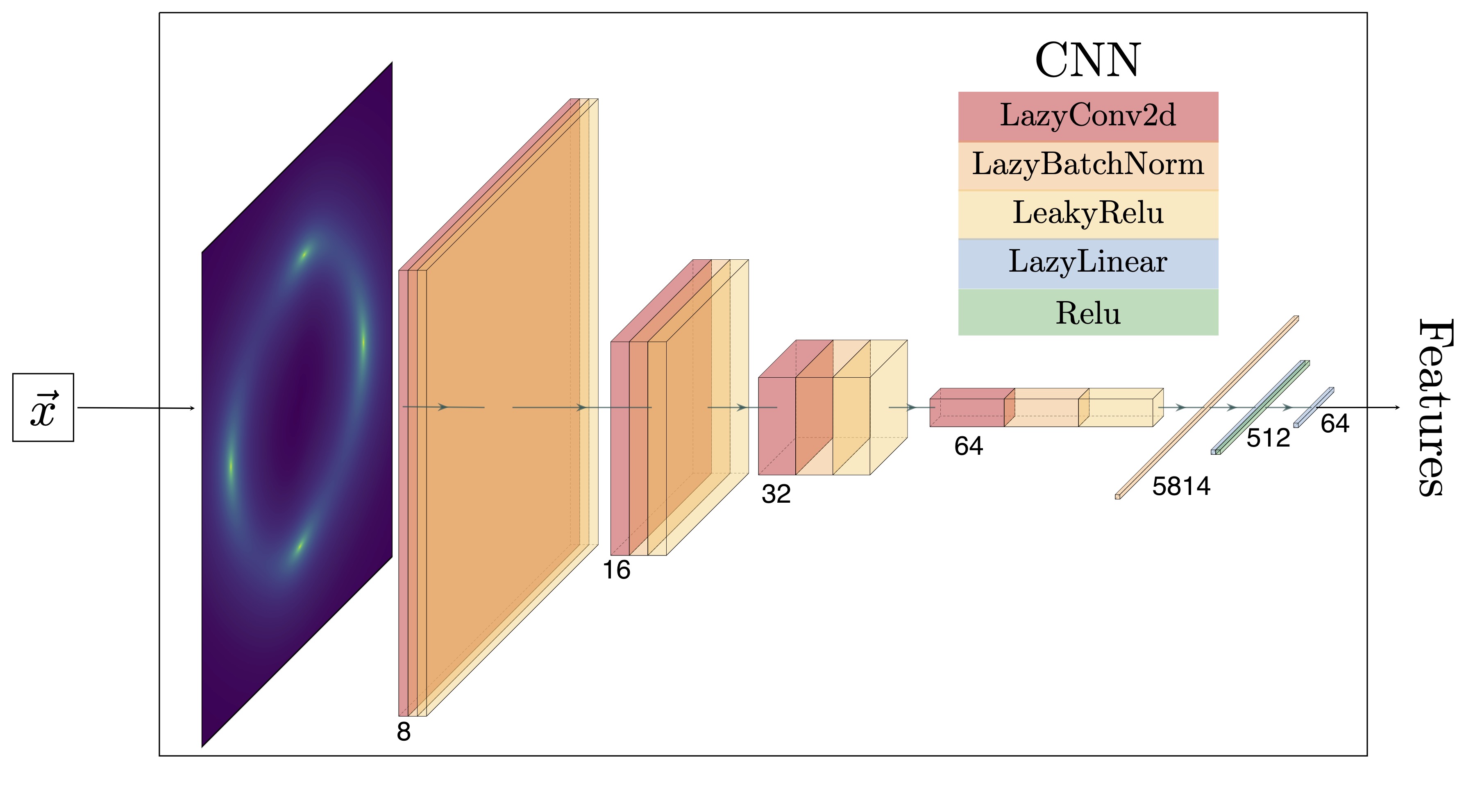}
\caption{Illustration of the embedding \gls*{cnn} architecture used in the first part of the pipeline to constrain lens and source parameters. The observation $\bx$ gets compressed into features: estimates of the best possible data summary statistic, by the \gls*{cnn}. In describing the \gls*{cnn} layers we follow \texttt{PyTorch} \citep{pytorch} convention. To create the illustration we have used \citet{PlotNeuralNet}.
}
\label{fig:cnn}
\end{figure}

For the embedding network, in both steps of the pipeline, we adopt a simple \gls*{cnn}. In \autoref{fig:cnn} we show the \gls*{cnn} architecture used to constrain lens and source parameters. The one used to estimate the cutoff mass has a similar structure.

\subsection{Constraining lens and source parameters}
\label{subsec:constrain}

\begin{figure*}
\centering
\includegraphics[width=\linewidth]{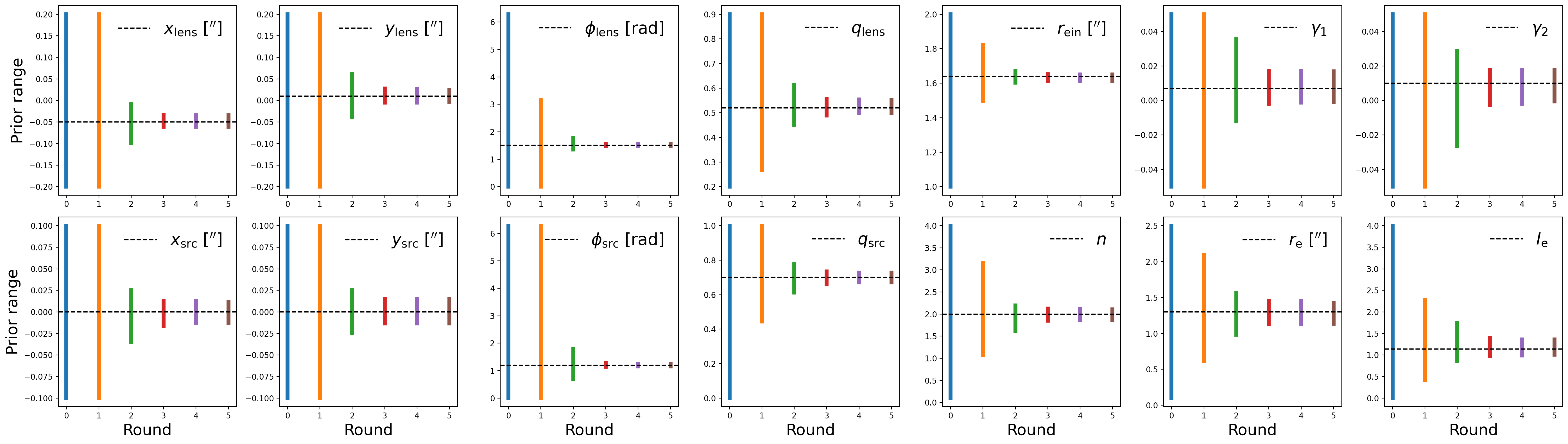}
\caption{Constrained proposal distribution. Visualization of the sequential truncation of the lens and source proposal distributions over the 6 rounds of training. The particular target is the first mock image (framed in orange in \autoref{fig:mock}), whose parameters are depicted as black dashed horizontal lines.
}
\label{fig:bounds}
\end{figure*}

\begin{figure}
\centering
\includegraphics[width=\linewidth]{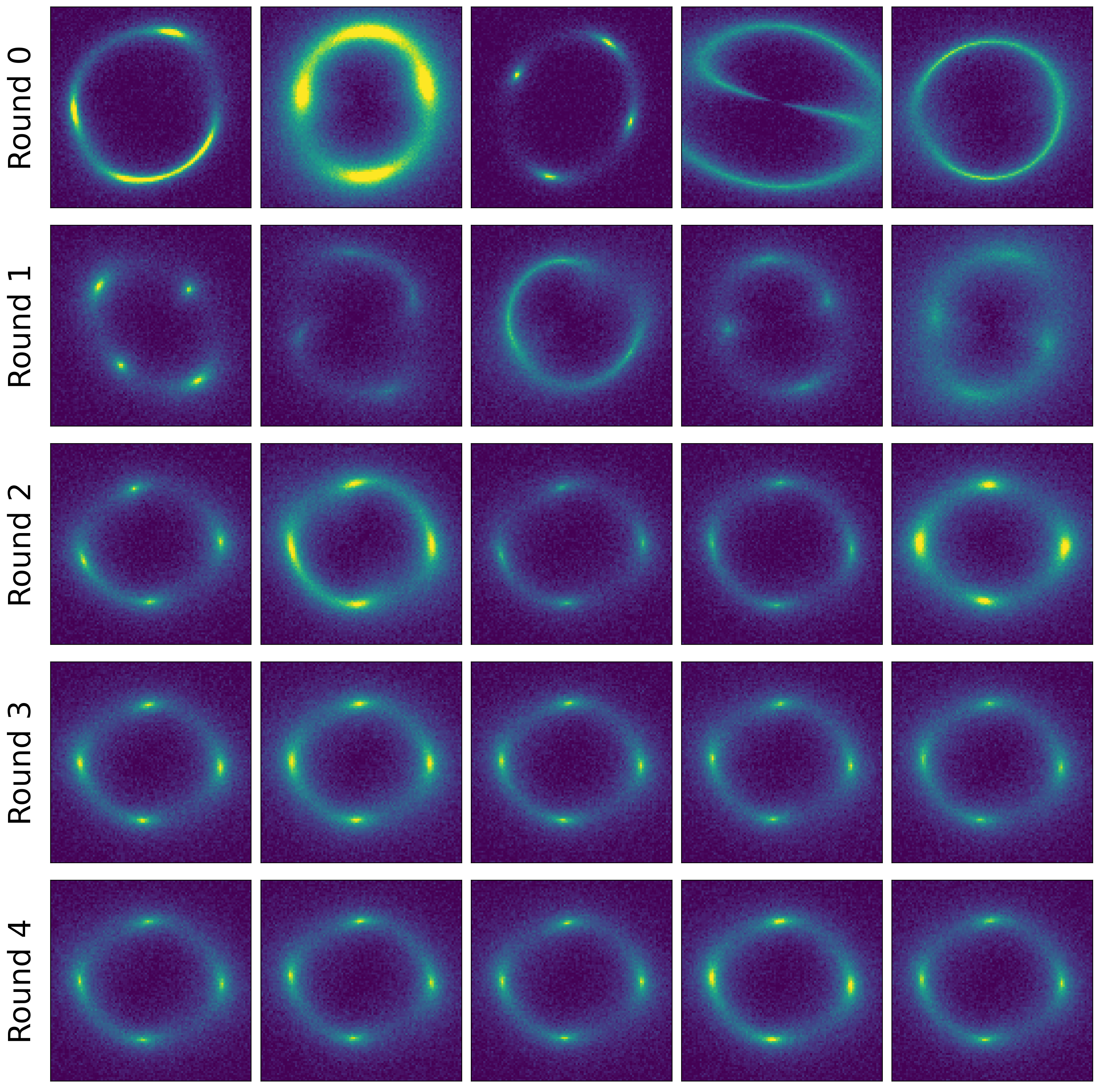}
\caption{Training data targeting the first mock observation (framed in orange in \autoref{fig:mock}). In each row, we show five examples of training data for the first five rounds. In the first round, we sample our data from the initial prior shown in \autoref{tab:model}. For the following rounds, the lens and source parameters are sampled from the constrained proposal distributions, obtained by evaluating the network trained with the previous round dataset on our target observation (see \autoref{subsec:constrain}). It is evident that with each round the training data more closely resembles the target image $\bx$
}
\label{fig:targeted_data}
\end{figure}

\begin{figure*}
\centering
\includegraphics[width=\linewidth]{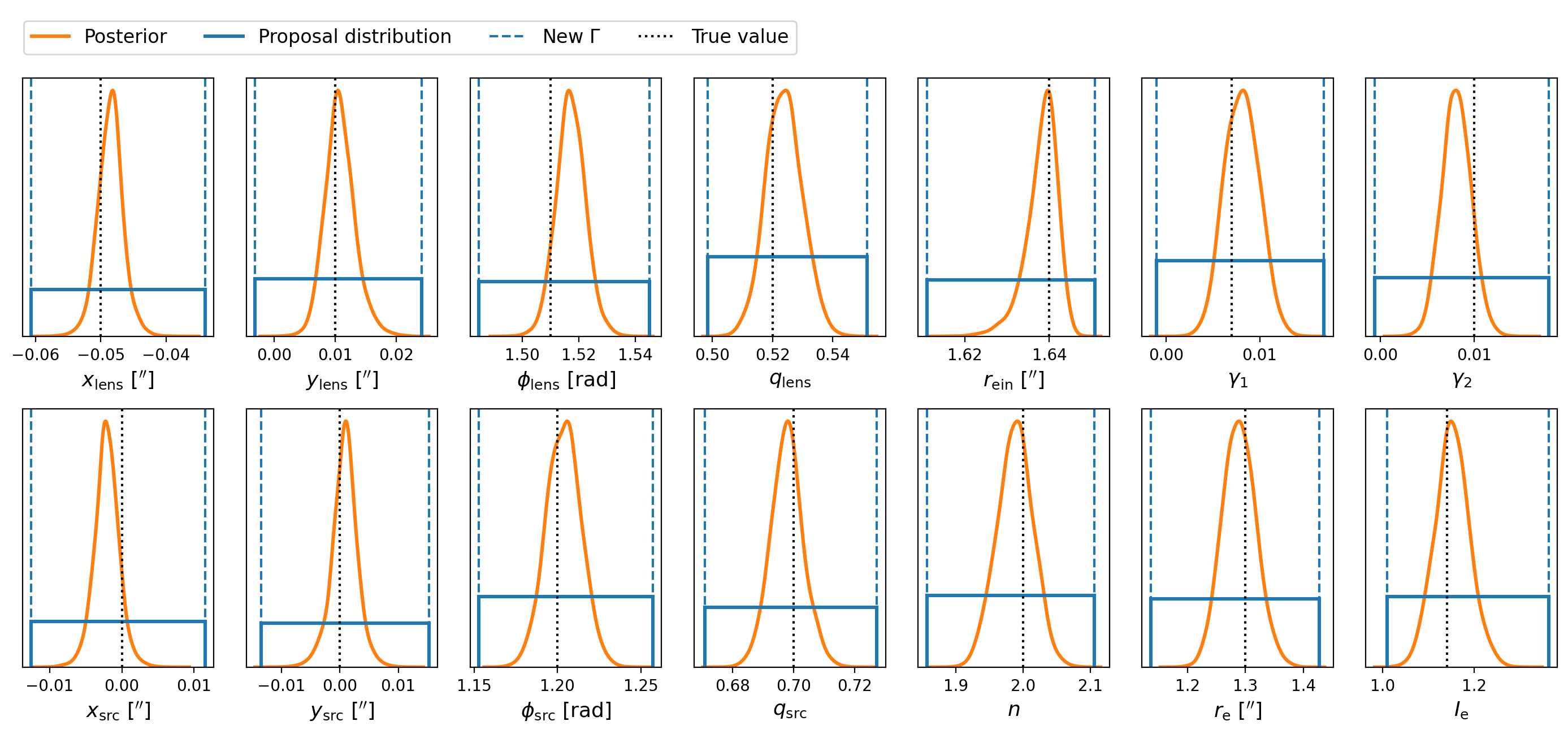}
\caption{Lens and source parameters posteriors. In solid blue we show the last round of constrained proposal distributions for the first (upper left corner, framed in orange) target image in \autoref{fig:mock}. The dotted black lines correspond to the true lens and source parameters values with which we have generated the target image. In orange, we show the estimated posteriors for lens and source parameters in the last training round. Based on the predetermined threshold, the new bounding limits $\Gamma$ (dashed blue) do not change significantly from the previous constrained proposal distribution region, so it is not possible to constrain the proposal distribution more and we stop the truncation procedure.
}
\label{fig:lens_source_post}
\end{figure*}

We constrain lens and source parameters regions with \gls*{tmnre}, as described in \autoref{subsec:tmre}, with multiple sampling and training rounds.

In total, we perform $6$ sampling and training rounds. In each round, we simulate $10^5$ observations, of which $90\%$ are used as the training dataset, and the remaining $10\%$ as the validation dataset. Evaluations of the network on the mock target image are used to truncate the  training data proposal distribution after each round, so that the region for lens and source parameters is targeted. The first training round is performed on the dataset generated from the initial source and lens parameters priors, shown in \autoref{tab:model}. In \autoref{fig:bounds} we show the initial prior and the following constrained proposal distributions. It can be seen that after the first round just a few of the parameters proposal distributions get truncated, e.g. the Einstein radius. By having truncated these initial parameters, in the following rounds the other parameters can be better learned by the network and so constrained. In \autoref{fig:targeted_data}, we show samples from the first five training datasets, which demonstrate that the constrained regions are indeed the ones that are likely to produce data similar to the targeted image $\bx$. After the sixth round of training, it is not possible anymore to truncate the proposal distribution region based on the predetermined threshold, as seen in \autoref{fig:lens_source_post}. The truncation scheme has then efficiently identified the constrained region for lens and source parameters consistent with the targeted observation. 

Using the last constrained dataset, it is then easier in the second step of the pipeline to train a marginal neural ratio estimator to perform the final inference on the cutoff mass, as explained in \autoref{subsec:tmre}. 

We would like to stress that these constrained proposal distributions correctly account for lens and source parameters uncertainties. In all our simulated data, the substructure parameters $\btheta_h$ are randomly sampled from their prior, in order to account for the presence of substructure. This has the desirable outcome of approximately accounting for the average effect that an additional mass component has on the main lens parameters (e.g.\ inferring an unbiased Einstein radius) and contributes to the source and lens uncertainties.

\subsection{Dark matter inference}
\label{subsec:dm}
For the second step of the pipeline, we train an inference network to learn the cutoff mass on the last constrained dataset. 

From initial tests, we have found that features from a single image are very hard to learn for the classifier, resulting in a very noisy ratio estimator.
In order to reduce the estimator uncertainty, we then train the cutoff mass classifier on a dataset $X^N=\{\bx_1, ..., \bx_N\}$ of $N$ different observations. For each observation, first, we constrain its lens and source parameters as explained in \autoref{subsec:constrain}. Then, we train the cutoff classifier on the concatenation of the features coming from their embedding networks, effectively learning $r(X^N,\vartheta)$. 
Note that the images in one dataset are sampled with the same cutoff mass $\mhm$, but different lens, source, and substructures realizations. In fact, our final goal is to apply the full pipeline to real data, which will all have different source, lens, and substructures configurations, but will have encoded the same \gls*{dm} properties. 

\begin{figure*}
\centering
\includegraphics[width=\linewidth]{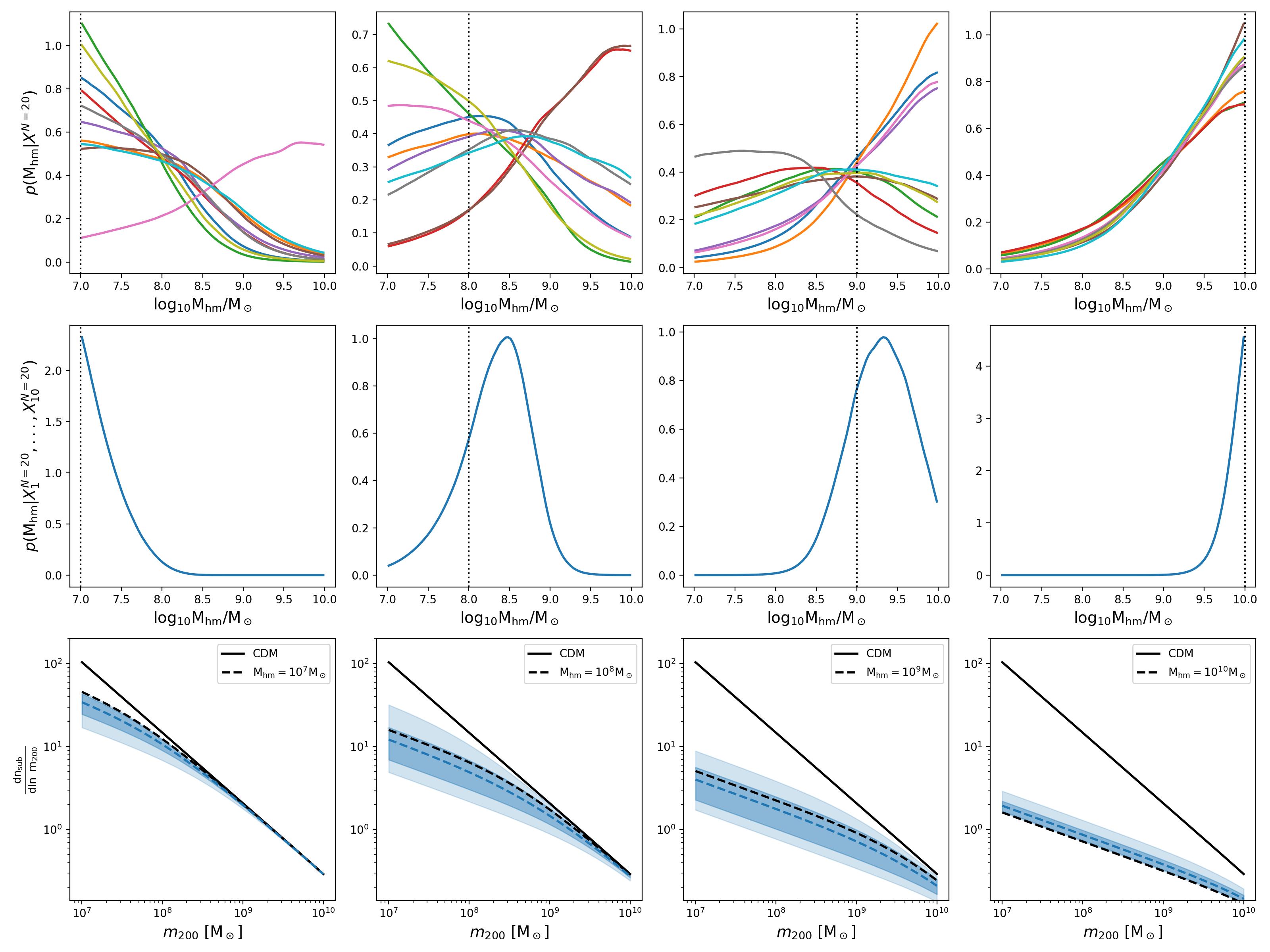}
\caption{\textit{Top:} Approximate posteriors for the half-mode mass derived from 10 different sets of 20 images. The dotted black line represents the true value of the half mode mass with which we have generated the images ($10^7, 10^8, 10^9,  10^{10} \ \si{\solmass}$). 
\textit{Middle:} We show the approximate posterior resulting from the combination of the $M=10$ different posteriors shown in the first column, as explained in the text (\autoref{subsec:dm}).
\textit{Bottom:} Subhalo mass function constraints derived from the cutoff mass posterior shown in the second column. The black solid line shows the \gls*{cdm} subhalo mass function according to \autoref{eq:shmf}, whereas the black dashed one shows the \gls*{wdm} subhalo mass function according to \autoref{eq:cutoff}, given the true cutoff mass shown in the label. The blue dashed line shows the mean of the \gls*{wdm} subhalo mass function obtained by sampling $1000$ samples from the cutoff mass posterior shown in the second panel and using this value in \autoref{eq:cutoff}. We also show the central \num{68} and \num{95} percentiles as shaded bands. These plots show how uncertain the subhalo mass function is under the assumption that it has the functional form in \autoref{eq:cutoff} with parameters from \citet{Lovell_2020}.
}
\label{fig:mhm_results}
\end{figure*}

In the first row of \autoref{fig:mhm_results} we show the results from the inference network on ten test sets of lenses generated with a $\mhm$ value of $10^7, 10^8, 10^9$ and, $10^{10} \ \si{\solmass}$. Each curve is the posterior obtained for a set of $N=20$ lenses. Each of the mock observations has lens and source parameters sampled from their own final constrained proposal distribution, and different substructure population.  

Now that we have reduced the estimator noise, it is straightforward to perform inference on a group of sets of images by combining their ratios. Given a dataset $X^N=\{\bx_1, ..., \bx_N\}$ of images, the combined ratio for multiple $M$ datasets is simply given by $r(X^N_M,\vartheta) \propto \prod_{i=1}^M r(X^N_i, \vartheta)$, where the proportionality is a ratio of evidences, independent of the parameter value, so it only accounts for a proper normalisation \citep{Brehmer_2019, Hermans_2020}.
In the second row of \autoref{fig:mhm_results} we show the results for the combination of the $M=10$ different posteriors shown in the first column.

In the third row we show a combined posterior for the \gls*{wdm} mass function from 200 images ($M=10$ sets of $N=20$ images). These plots show the uncertainty in the subhalo mass function under the assumption that it has the functional form in \autoref{eq:cutoff} with parameters from \citet{Lovell_2020}.

These first results show that our method is sensitive to the low-mass end of the \gls*{hmf}, and that we have unbiased results from combining just 10 sets of 20 observations, given that in the second panel of \autoref{fig:mhm_results} the true input value for the half-mode mass $\mhm$ is consistently contained within the estimated posterior. In \autoref{subsec:test} we will show a more sophisticated method to assess the statistical behaviour of
our inference results.

\begin{figure*}
\centering
\includegraphics[width=\linewidth]{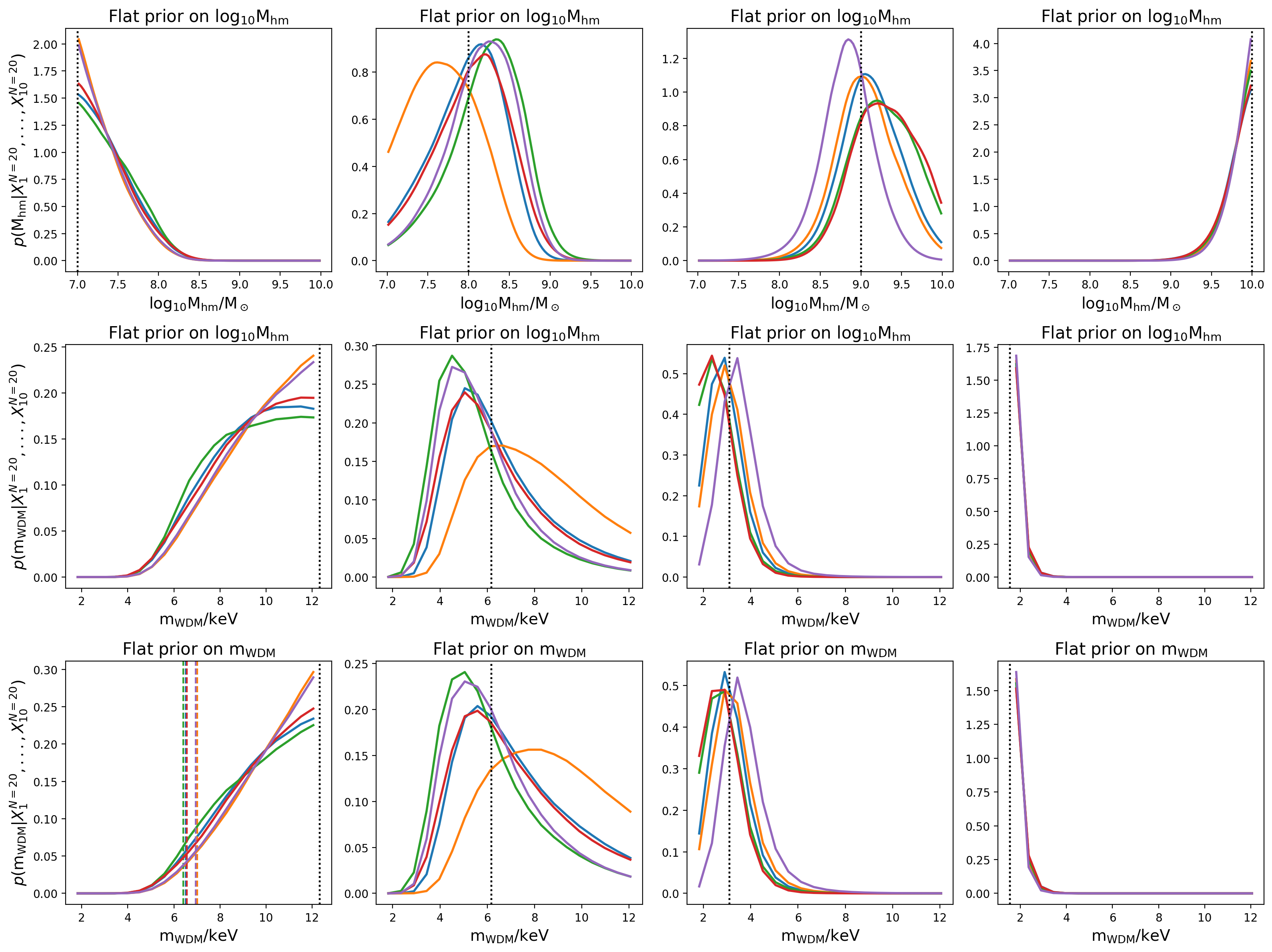}
\caption{\textit{Top:} We show five examples of combined posterior of $M=10$ sets of $N=20$ observations in terms of the cutoff mass (as the second row in \autoref{fig:mhm_results}). The dotted black line represents the true input value of the half mode mass with which we have generated the analyzed mock observations ($10^7, 10^8, 10^9,  10^{10} \ \si{\solmass}$). 
\textit{Middle:} Same results as shown in the first column but for the \gls*{wdm} mass. The dotted black line represents the true value of the WDM mass with which we have generated the analyzed mock observations, given the mapping between \gls*{dm} cutoff and \gls*{dm} mass in \autoref{subsec:free-streaming}. The WDM mass posteriors assume a flat prior on the cutoff mass.
\textit{Bottom:} Same results as shown in the first column but for the \gls*{wdm} mass and assuming a flat prior on the latter. In the first plot of the row, we show for the five examples the expected 95\% credible lower limit on the WDM mass for the highest value of our prior distribution. 
}
\label{fig:wdm_results}
\end{figure*}

Furthermore, we can translate the constraints we obtain on the cutoff mass to constraints on the WDM mass given the mapping between those two quantities defined in \autoref{subsec:free-streaming}. In \autoref{fig:wdm_results} we show  our results for the \gls*{wdm} mass. Each column corresponds to a different cutoff mass input value, so a different WDM mass. In the first row, we plot five examples of the combined posterior density for $\log_{10}\mhm$ of $M=10$ sets of $N=20$ observations. In the second row, we show the corresponding color-coded five examples for $m_{\mathrm{WDM}}$. In this case, we just transform the posterior from the first row using the parameterisation shown in \autoref{subsec:free-streaming}, so we assume a flat prior on $\log_{10}\mhm$. Finally, in the last row, we show the WDM mass posterior densities assuming a flat prior on the latter. The posteriors in the second and third row are not actually the same because a flat prior $\log_{10}\mhm$ is different from a flat prior on $m_{\mathrm{WDM}}$.

\subsection{Credible interval testing}
\label{subsec:test}

\begin{figure}
    \centering
    \includegraphics[width=\linewidth]{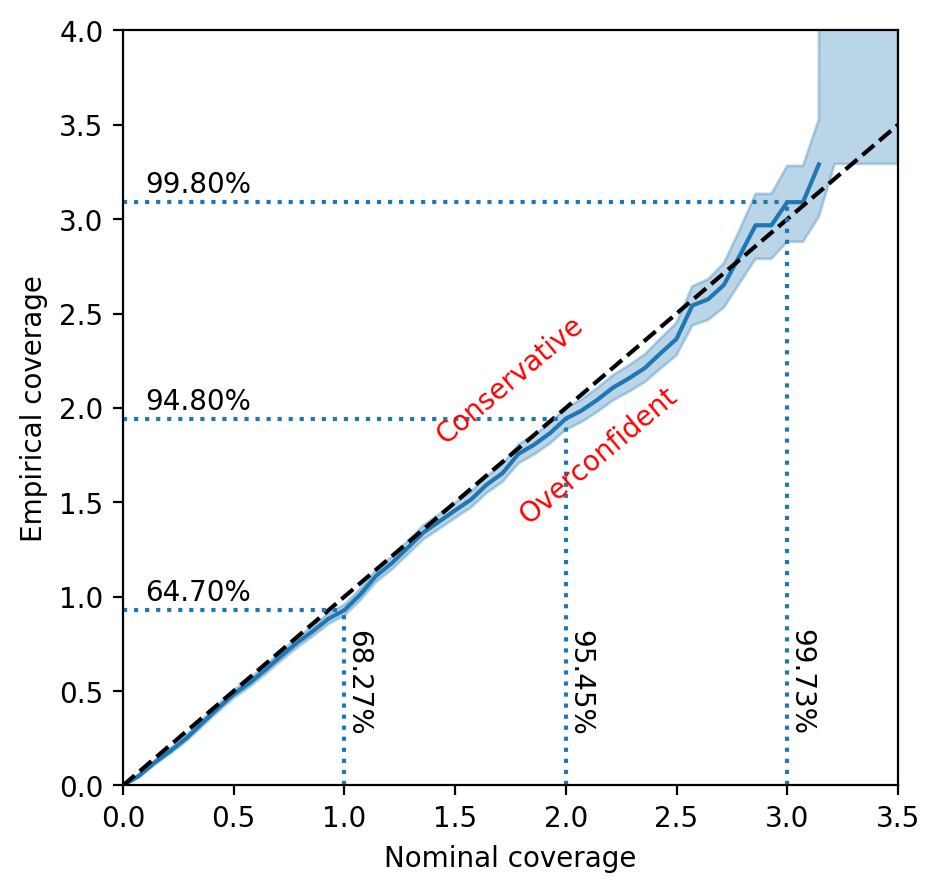}
    \caption{Empirical versus nominal expected coverage probabilities for the cutoff mass inference network. In case the line lies above (below) the black dashed diagonal line, the credible intervals are conservative (overconfident) and contain the true value with a frequency higher (lower) than nominally expected. We show the empirical (nominal) probabilities as horizontal (vertical) text.
}
\label{fig:coverage}
\end{figure}

We would like to directly test and validate the statistical behaviour of our inference results by determining the expected coverage of the ratio estimator produced by the network. This can be easily done in \swyft thanks to local amortization \citep{Miller_2021}.
The goal is to compare the nominal and empirical expected coverage probabilities of estimated Bayesian credible intervals, which should coincide for a well-calibrated estimator. For the statistical formalism and definition of credible region and expected coverage probability, we refer the reader to \cite{Hermans_2021}. In brief, an ideal estimator has matching empirical and nominal expected coverage, a conservative one predicts lower credibility than empirically obtained, and an overconfident one has higher nominal than empirical credibility. In plots like \autoref{fig:coverage}, the line for an ideal ratio estimator should perfectly align with the diagonal, whereas for a conservative (overconfident) estimator, it will lie above (below) the diagonal.
In combination with visually checking the posteriors, this test supports the accuracy of the posterior estimator and is also particularly useful when one does not have access to the ground truth against which to compare the results.
In \autoref{fig:coverage} we show the empirical versus nominal expected coverage probabilities for the cutoff mass inference network. We can see that the inference network for the half-mode mass has converged with good expected coverage. 

\section{Discussion}
\label{sec:discussion}
\hl
In this section, we discuss the improvements to the model and inference question which need to be addressed before we can safely apply our pipeline to the analysis of real data. 

First, we have neglected effects such as inadequate lens light subtraction and assumed the lens light to be known. 
Regarding the noise model, we did not account for correlated pixel noise due to instrumental effects including the telescope's PSF (e.g. see \citealt{Wagner-Carena_2022}).

In this work, we have employed an analytic parameterisation (the Sérsic profile) as a lensed source light distribution model, which is adequate to analyze low-resolution images. However, to accurately model higher-fidelity lensing observations, such as those from on-going (e.g. HST) and future (e.g. JWST, ELT, SKA) telescopes, more complex source models need to be employed. Existing models, in order of complexity, are regularised pixellation of the source plane (see, e.g., \citealt{Suyu_2006, Karchev_2021, Vegetti_2009a}), source modelling through basis functions (e.g. shapelets \citep{Birrer_2018} or wavelets \citep{Galan_2021}) attached to the source plane, and deep learning approaches (see, e.g., \citealt{Adam_2022, Morningstar_2019}).
The ability to accurately and precisely reconstruct the complex morphology of strong-lensing sources is of the utmost importance, as to disentangle the source surface brightness inhomogeneities from the percent-level fluctuations introduced by substructures in the lens. We anticipate that using sources with more complex morphologies will result in higher sensitivity to the DM cutoff mass, provided that it is possible to model these sources.
In fact, the residuals between the image of an extended source lensed by the total lens potential (accounting for substructures), and that of the same source lensed only by the main lens component are proportional to the gradient of that source evaluated in the image plane \citep[Equation 16]{Cyr-Racine_2019}. 

Regarding DM modelling, validation of our smoothing scheme (\autoref{subsubsec:smoothing}) is required to accurately account for DM free-streaming effects. Moreover, we should account for uncertainties due to the assumed halo density profile by considering different DM distributions around galaxies (see, e.g., \citealt{Salucci_2019} for a review).
 
Finally, we would like to draw the reader's attention on the fact that in our modeling we assume that the halo mass of the lens is known exactly from its Einstein radius (see \autoref{subsubsec:sub-dist}). 
This is a strong assumption that has as a consequence the separation of substructure parameters $\btheta_h$ and lens parameter $\btheta_l$ once we marginalize the posterior probability over the halo mass in \autoref{eq:model}.
The inference question we have addressed in this work, constraining the cutoff mass of the subhalo mass distribution, is then a simplified version of the real one, which is to simultaneously determine the halo mass and subhalo mass distribution of the lenses from real data (see, e.g., \citealt{Birrer_2017}). 

We believe there are no major obstacles in incorporating all of these modeling components in our framework without fundamentally altering the inference procedure.

\section{Conclusions}
\label{conclusions}

Strong gravitational lensing as a probe of the particle nature of \gls*{dm} has sparked much interest over the last few years. Moreover, the development of fast and accurate techniques to extract information from strong lensing images is well motivated by the wealth of new high-resolution strong lensing observations that will become available in the near future.

In this work, we have presented the first step towards a new neural simulation based inference pipeline (see \autoref{analysis}) to analyse present and future strong gravitational lensing systems in order to constrain the cutoff in the \gls*{dm} \gls*{hmf}, and so the \gls*{dm} mass. To this end, we have used a recent machine learning development, \gls*{tmnre}, that makes it possible to \textit{target} the analysis to a specific observation rather than amortize over all possible variations in lensing systems, making inference more efficient and precise.  Thanks to \gls*{tmnre}, we overcome the computational challenges of traditional \gls*{mcmc}, nested sampling and trans-dimensional \gls*{mcmc} methods, by directly learning the marginal posterior for the parameter of interest from the observation. \gls*{tmnre} leverages neural networks to directly learn the best summary statistic possible from the full input data, without having to compress the observation into hand-crafted summary statistics. This work is then a step forward towards making the analysis of strong lensing images for \gls*{dm} science faster, more efficient, and more accurate. In addition, our inference results can be validated with expected coverage tests (see \autoref{subsec:test}). 

Our key results can be summarized as follows:
\begin{itemize}
    \item Thanks to our targeted approach, we are able to correctly estimate the lens and source parameters uncertainties, accounting for the presence of substructures in the mass range $[10^7,\ 10^{10}]\ \si{\solmass}$. We use the final lens and source parameters truncated proposal distributions (see \autoref{subsec:constrain}) to generate a targeted training dataset in order to infer the \gls*{dm} cutoff. 
    \item In the case that \gls*{dm} is warm, we are able to infer the location of the cutoff in the \gls*{hmf} in the $[10^7,\ 10^{10}]\ \si{\solmass}$ mass range by combining up to 200 observations (see \autoref{subsec:dm}). We show our results in \autoref{fig:mhm_results}. By construction, these results are correctly marginalized over model uncertainties and have proper expected coverage (see \autoref{subsec:test}).
    \item A cutoff mass posterior translates into a posterior on the \gls*{wdm} mass, given the mapping in \autoref{subsec:free-streaming}. We show our results in \autoref{fig:wdm_results} for a flat prior on the cutoff mass and a flat prior on the \gls*{wdm} mass. We obtain an expected 95\% credible lower limits around 6.5 $\si{\keV}$ in the case of the scenario closest to CDM (see the bottom left panel in \autoref{fig:wdm_results}), given the adopted prior and the various assumptions of our simulation model that will be discussed below.
\end{itemize}

Throughout this study, we have made a number of simplifying assumptions for the, halo mass of the lens, source light profile, and substructure models. We have also neglected effects such as inadequate lens light subtraction, realistic \gls*{psf} modeling, and correlated pixel noise due to effects including the telescope's \gls*{psf}. Before this analysis pipeline can be safely extended to real observations, these assumptions need to be correctly addressed, as discussed in \autoref{sec:discussion}. 

In this work, we have demonstrated that, in principle, the \gls*{dm} cutoff mass signal can be statistically extracted from a population of small-scale dark matter halos by a neural network using \gls*{tmnre}.
In future works, we plan on studying the correlation between different subhalo mass function parameters (e.g. its normalization, slope, and cutoff mass), and the one between the halo mass and subhalo mass distribution of the lenses, using, on one hand, more advanced modeling techniques (as specified above) on multi-band observations, and, on the other, better neural network architectures to target low \gls*{snr} scenarios. 

Finally, we note that, thanks to its flexibility, our pipeline can incorporate any arbitrary \gls*{dm} model, as long as it specifies the form of the \gls*{hmf} and the density profiles of individual substructures. We are optimistic that the presented Bayesian inference pipeline will be able to constrain the amount of substructures, pinning down \gls*{dm} nature, using both the strong lensing images that exist today and the wealth of new strong lensing data coming from near-future observatories.

\section*{Acknowledgements}
We thank Benjamin Kurt Miller and Elias Dubbeldam for helpful discussions. We thank the anonymous referee for a careful reading and helpful comments.

This work is part of a project that has received funding from the European Research Council (ERC) under the European Union’s Horizon 2020 research and innovation program (Grant agreement No. 864035 -- UnDark).

A.C. received funding from the Netherlands eScience Center (grant number ETEC.2019.018) and the Schmidt Futures Foundation.
C.C. acknowledges the support of the Dutch Research Council (NWO Veni 192.020).

This work was carried out on the Lisa Compute Cluster at SURFsara. We acknowledge the use of the \texttt{python} \citep{python} modules, \texttt{matplotlib} \citep{matplotlib}, \texttt{seaborn} \citep{seaborn}, \texttt{numpy} \citep{numpy},  \texttt{scipy} \citep{scipy}, \texttt{AstroPy} \citep{astropy}, \texttt{PyTorch} \citep{pytorch}, \texttt{Pyro} \citep{pyro}, \texttt{tqdm} \citep{tqdm}, and \texttt{jupyter} \citep{jupyter}.

\section*{Data Availability}
The data underlying this article will be shared on reasonable request to the corresponding author.

\bibliographystyle{mnras}
\bibliography{references} 

\begin{thebibliography}{}
\makeatletter
\relax
\def\mn@urlcharsother{\let\do\@makeother \do\$\do\&\do\#\do\^\do\_\do\%\do\~}
\def\mn@doi{\begingroup\mn@urlcharsother \@ifnextchar [ {\mn@doi@}
  {\mn@doi@[]}}
\def\mn@doi@[#1]#2{\def\@tempa{#1}\ifx\@tempa\@empty \href
  {http://dx.doi.org/#2} {doi:#2}\else \href {http://dx.doi.org/#2} {#1}\fi
  \endgroup}
\def\mn@eprint#1#2{\mn@eprint@#1:#2::\@nil}
\def\mn@eprint@arXiv#1{\href {http://arxiv.org/abs/#1} {{\tt arXiv:#1}}}
\def\mn@eprint@dblp#1{\href {http://dblp.uni-trier.de/rec/bibtex/#1.xml}
  {dblp:#1}}
\def\mn@eprint@#1:#2:#3:#4\@nil{\def\@tempa {#1}\def\@tempb {#2}\def\@tempc
  {#3}\ifx \@tempc \@empty \let \@tempc \@tempb \let \@tempb \@tempa \fi \ifx
  \@tempb \@empty \def\@tempb {arXiv}\fi \@ifundefined
  {mn@eprint@\@tempb}{\@tempb:\@tempc}{\expandafter \expandafter \csname
  mn@eprint@\@tempb\endcsname \expandafter{\@tempc}}}

\bibitem[\protect\citeauthoryear{Abolfathi et~al.,}{Abolfathi
  et~al.}{2021}]{LSST}
Abolfathi B.,  et~al., 2021, \mn@doi [The Astrophysical Journal Supplement
  Series] {10.3847/1538-4365/abd62c}, 253, 31

\bibitem[\protect\citeauthoryear{Adam, Perreault-Levasseur  \& Hezaveh}{Adam
  et~al.}{2022}]{Adam_2022}
Adam A.,  Perreault-Levasseur L.,   Hezaveh Y.,  2022, Pixelated Reconstruction
  of Gravitational Lenses using Recurrent Inference Machines (\mn@eprint
  {arXiv} {2207.01073})

\bibitem[\protect\citeauthoryear{Ade et~al.,}{Ade et~al.}{2016}]{Planck_2015}
Ade P. A.~R.,  et~al., 2016, \mn@doi [Astronomy & Astrophysics]
  {10.1051/0004-6361/201525830}, 594, A13

\bibitem[\protect\citeauthoryear{Alexander, Gleyzer, McDonough, Toomey  \&
  Usai}{Alexander et~al.}{2020}]{Alexander_2020}
Alexander S.,  Gleyzer S.,  McDonough E.,  Toomey M.~W.,   Usai E.,  2020,
  \mn@doi [The Astrophysical Journal] {10.3847/1538-4357/ab7925}, 893, 15

\bibitem[\protect\citeauthoryear{Amorisco et~al.,}{Amorisco
  et~al.}{2021}]{Amorisco_2021}
Amorisco N.~C.,  et~al., 2021, \mn@doi [Monthly Notices of the Royal
  Astronomical Society] {10.1093/mnras/stab3527}, 510, 2464

\bibitem[\protect\citeauthoryear{{Astropy Collaboration} et~al.,}{{Astropy
  Collaboration} et~al.}{2018}]{astropy}
{Astropy Collaboration} et~al., 2018, \mn@doi [\aj] {10.3847/1538-3881/aabc4f},
  \href {https://ui.adsabs.harvard.edu/abs/2018AJ....156..123A} {156, 123}

\bibitem[\protect\citeauthoryear{Baltz, Marshall  \& Oguri}{Baltz
  et~al.}{2009}]{Baltz_2009}
Baltz E.~A.,  Marshall P.,   Oguri M.,  2009, \mn@doi [Journal of Cosmology and
  Astroparticle Physics] {10.1088/1475-7516/2009/01/015}, 2009, 015–015

\bibitem[\protect\citeauthoryear{Bayer, Chatterjee, Koopmans, Vegetti, McKean,
  Treu  \& Fassnacht}{Bayer et~al.}{2018}]{Bayer_2018}
Bayer D.,  Chatterjee S.,  Koopmans L. V.~E.,  Vegetti S.,  McKean J.~P.,  Treu
  T.,   Fassnacht C.~D.,  2018, Observational constraints on the sub-galactic
  matter-power spectrum from galaxy-galaxy strong gravitational lensing
  (\mn@eprint {arXiv} {1803.05952})

\bibitem[\protect\citeauthoryear{Bingham et~al.,}{Bingham et~al.}{2019}]{pyro}
Bingham E.,  et~al., 2019, J. Mach. Learn. Res., 20, 28:1

\bibitem[\protect\citeauthoryear{Birrer \& Amara}{Birrer \&
  Amara}{2018}]{Birrer_2018}
Birrer S.,  Amara A.,  2018, \mn@doi [Physics of the Dark Universe]
  {10.1016/j.dark.2018.11.002}, 22, 189

\bibitem[\protect\citeauthoryear{Birrer, Amara  \& Refregier}{Birrer
  et~al.}{2017}]{Birrer_2017}
Birrer S.,  Amara A.,   Refregier A.,  2017, \mn@doi [Journal of Cosmology and
  Astroparticle Physics] {10.1088/1475-7516/2017/05/037}, 2017, 037–037

\bibitem[\protect\citeauthoryear{Bolton, Burles, Koopmans, Treu  \&
  Moustakas}{Bolton et~al.}{2006}]{SLACS}
Bolton A.~S.,  Burles S.,  Koopmans L. V.~E.,  Treu T.,   Moustakas L.~A.,
  2006, \mn@doi [The Astrophysical Journal] {10.1086/498884}, 638, 703–724

\bibitem[\protect\citeauthoryear{Bond, Szalay  \& Turner}{Bond
  et~al.}{1982}]{Bond_1982}
Bond J.~R.,  Szalay A.~S.,   Turner M.~S.,  1982, \mn@doi [Phys. Rev. Lett.]
  {10.1103/PhysRevLett.48.1636}, 48, 1636

\bibitem[\protect\citeauthoryear{Boyarsky, Drewes, Lasserre, Mertens  \&
  Ruchayskiy}{Boyarsky et~al.}{2019}]{Boyarsky_2019}
Boyarsky A.,  Drewes M.,  Lasserre T.,  Mertens S.,   Ruchayskiy O.,  2019,
  \mn@doi [Progress in Particle and Nuclear Physics]
  {10.1016/j.ppnp.2018.07.004}, 104, 1–45

\bibitem[\protect\citeauthoryear{Brehmer, Mishra-Sharma, Hermans, Louppe  \&
  Cranmer}{Brehmer et~al.}{2019}]{Brehmer_2019}
Brehmer J.,  Mishra-Sharma S.,  Hermans J.,  Louppe G.,   Cranmer K.,  2019,
  \mn@doi [The Astrophysical Journal] {10.3847/1538-4357/ab4c41}, 886, 49

\bibitem[\protect\citeauthoryear{Brennan, Benson, Cyr-Racine, Keeton, Moustakas
   \& Pullen}{Brennan et~al.}{2019}]{Brennan_2019}
Brennan S.,  Benson A.~J.,  Cyr-Racine F.-Y.,  Keeton C.~R.,  Moustakas L.~A.,
   Pullen A.~R.,  2019, \mn@doi [Monthly Notices of the Royal Astronomical
  Society] {10.1093/mnras/stz1607}, 488, 5085–5092

\bibitem[\protect\citeauthoryear{Brewer, Huijser  \& Lewis}{Brewer
  et~al.}{2015}]{Brewer_2015}
Brewer B.~J.,  Huijser D.,   Lewis G.~F.,  2015, Trans-Dimensional Bayesian
  Inference for Gravitational Lens Substructures (\mn@eprint {arXiv}
  {1508.00662})

\bibitem[\protect\citeauthoryear{Brownstein et~al.,}{Brownstein
  et~al.}{2011}]{BELLS}
Brownstein J.~R.,  et~al., 2011, \mn@doi [The Astrophysical Journal]
  {10.1088/0004-637x/744/1/41}, 744, 41

\bibitem[\protect\citeauthoryear{Bullock}{Bullock}{2010}]{Bullock_2010}
Bullock J.~S.,  2010, Notes on the Missing Satellites Problem,
  \mn@doi{10.48550/ARXIV.1009.4505}, \url {https://arxiv.org/abs/1009.4505}

\bibitem[\protect\citeauthoryear{{\c{C}}a\v{g}an~\c{S}eng\"{u}l, Tsang,
  Diaz~Rivero, Dvorkin, Zhu  \& Seljak}{{\c{C}}a\v{g}an~\c{S}eng\"{u}l
  et~al.}{2020}]{Sengul_2020}
{\c{C}}a\v{g}an~\c{S}eng\"{u}l A.,  Tsang A.,  Diaz~Rivero A.,  Dvorkin C.,
  Zhu H.-M.,   Seljak U.,  2020, \mn@doi [Physical Review D]
  {10.1103/physrevd.102.063502}, 102

\bibitem[\protect\citeauthoryear{Chatterjee \& Koopmans}{Chatterjee \&
  Koopmans}{2017}]{Chatterjee_2017}
Chatterjee S.,  Koopmans L. V.~E.,  2017, \mn@doi [Monthly Notices of the Royal
  Astronomical Society] {10.1093/mnras/stx2674}, 474, 1762–1772

\bibitem[\protect\citeauthoryear{Ciotti \& Bertin}{Ciotti \&
  Bertin}{1999}]{Ciotti_1999}
Ciotti L.,  Bertin G.,  1999, Analytical properties of the $R^{1/m}$ luminosity
  law (\mn@eprint {arXiv} {astro-ph/9911078})

\bibitem[\protect\citeauthoryear{Cole, Miller, Witte, Cai, Grootes, Nattino  \&
  Weniger}{Cole et~al.}{2021}]{Cole_2021}
Cole A.,  Miller B.~K.,  Witte S.~J.,  Cai M.~X.,  Grootes M.~W.,  Nattino F.,
   Weniger C.,  2021, Fast and Credible Likelihood-Free Cosmology with
  Truncated Marginal Neural Ratio Estimation (\mn@eprint {arXiv} {2111.08030})

\bibitem[\protect\citeauthoryear{Colin, Avila‐Reese  \& Valenzuela}{Colin
  et~al.}{2000}]{Colin_2000}
Colin P.,  Avila‐Reese V.,   Valenzuela O.,  2000, \mn@doi [The Astrophysical
  Journal] {10.1086/317057}, 542, 622–630

\bibitem[\protect\citeauthoryear{Coogan, Karchev  \& Weniger}{Coogan
  et~al.}{2020}]{Coogan_2020}
Coogan A.,  Karchev K.,   Weniger C.,  2020, Targeted Likelihood-Free Inference
  of Dark Matter Substructure in Strongly-Lensed Galaxies (\mn@eprint {arXiv}
  {2010.07032})

\bibitem[\protect\citeauthoryear{Coogan, Correa, Karchev, Anau~Montel  \&
  Weniger}{Coogan et~al.}{2022}]{Coogan_2022}
Coogan A.,  Correa C.,  Karchev K.,  Anau~Montel N.,   Weniger C.,  2022,
  Characterizing subhalos in strong gravitational lenses with truncated
  marginal neural ratio estimation

\bibitem[\protect\citeauthoryear{Costa-Luis et~al.,}{Costa-Luis
  et~al.}{2021}]{tqdm}
Costa-Luis C.~d.,  et~al., 2021, {tqdm: A fast, Extensible Progress Bar for
  Python and CLI}, \mn@doi{10.5281/zenodo.5517697}, \url
  {https://doi.org/10.5281/zenodo.5517697}

\bibitem[\protect\citeauthoryear{Cranmer, Brehmer  \& Louppe}{Cranmer
  et~al.}{2020}]{Cranmer_2020}
Cranmer K.,  Brehmer J.,   Louppe G.,  2020, The frontier of simulation-based
  inference (\mn@eprint {arXiv} {1911.01429})

\bibitem[\protect\citeauthoryear{Cyr-Racine, Keeton  \& Moustakas}{Cyr-Racine
  et~al.}{2019}]{Cyr-Racine_2019}
Cyr-Racine F.-Y.,  Keeton C.~R.,   Moustakas L.~A.,  2019, \mn@doi [Physical
  Review D] {10.1103/physrevd.100.023013}, 100

\bibitem[\protect\citeauthoryear{Dalal \& Kochanek}{Dalal \&
  Kochanek}{2002}]{Dalal_2002}
Dalal N.,  Kochanek C.~S.,  2002, \mn@doi [The Astrophysical Journal]
  {10.1086/340303}, 572, 25–33

\bibitem[\protect\citeauthoryear{Daylan, Cyr-Racine, Rivero, Dvorkin  \&
  Finkbeiner}{Daylan et~al.}{2018}]{Daylan_2018}
Daylan T.,  Cyr-Racine F.-Y.,  Rivero A.~D.,  Dvorkin C.,   Finkbeiner D.~P.,
  2018, \mn@doi [The Astrophysical Journal] {10.3847/1538-4357/aaaa1e}, 854,
  141

\bibitem[\protect\citeauthoryear{Despali \& Vegetti}{Despali \&
  Vegetti}{2017}]{Despali_2017}
Despali G.,  Vegetti S.,  2017, \mn@doi [Monthly Notices of the Royal
  Astronomical Society] {10.1093/mnras/stx966}, 469, 1997–2010

\bibitem[\protect\citeauthoryear{Despali, Vegetti, White, Giocoli  \& van~den
  Bosch}{Despali et~al.}{2018}]{Despali_2018}
Despali G.,  Vegetti S.,  White S. D.~M.,  Giocoli C.,   van~den Bosch F.~C.,
  2018, \mn@doi [Monthly Notices of the Royal Astronomical Society]
  {10.1093/mnras/sty159}, 475, 5424–5442

\bibitem[\protect\citeauthoryear{Diaz~Rivero \& Dvorkin}{Diaz~Rivero \&
  Dvorkin}{2020}]{DiazRivero_2020}
Diaz~Rivero A.,  Dvorkin C.,  2020, \mn@doi [Physical Review D]
  {10.1103/physrevd.101.023515}, 101

\bibitem[\protect\citeauthoryear{Diaz~Rivero, Cyr-Racine  \&
  Dvorkin}{Diaz~Rivero et~al.}{2018}]{DiazRivero_2018a}
Diaz~Rivero A.,  Cyr-Racine F.-Y.,   Dvorkin C.,  2018, \mn@doi [Physical
  Review D] {10.1103/physrevd.97.023001}, 97

\bibitem[\protect\citeauthoryear{Drlica-Wagner et~al.,}{Drlica-Wagner
  et~al.}{2019}]{Drlica-Wagner_2019}
Drlica-Wagner A.,  et~al., 2019, Probing the Fundamental Nature of Dark Matter
  with the Large Synoptic Survey Telescope (\mn@eprint {arXiv} {1902.01055})

\bibitem[\protect\citeauthoryear{Díaz~Rivero, Dvorkin, Cyr-Racine, Zavala  \&
  Vogelsberger}{Díaz~Rivero et~al.}{2018}]{DiazRivero_2018b}
Díaz~Rivero A.,  Dvorkin C.,  Cyr-Racine F.-Y.,  Zavala J.,   Vogelsberger M.,
   2018, \mn@doi [Physical Review D] {10.1103/physrevd.98.103517}, 98

\bibitem[\protect\citeauthoryear{Galan, Peel, Joseph, Courbin  \& Starck}{Galan
  et~al.}{2021}]{Galan_2021}
Galan A.,  Peel A.,  Joseph R.,  Courbin F.,   Starck J.-L.,  2021, \mn@doi
  [Astronomy & Astrophysics] {10.1051/0004-6361/202039363}, 647, A176

\bibitem[\protect\citeauthoryear{Gardner et~al.,}{Gardner et~al.}{2006}]{JWST}
Gardner J.~P.,  et~al., 2006, \mn@doi [Space Science Reviews]
  {10.1007/s11214-006-8315-7}, 123, 485–606

\bibitem[\protect\citeauthoryear{{Gennaro}}{{Gennaro}}{2018}]{HST}
{Gennaro} M.,  2018, in , Vol.~4, WFC3 Data Handbook v. 4.
p.~4

\bibitem[\protect\citeauthoryear{Gilman, Birrer, Treu, Keeton  \&
  Nierenberg}{Gilman et~al.}{2018}]{Gilman_2018}
Gilman D.,  Birrer S.,  Treu T.,  Keeton C.~R.,   Nierenberg A.,  2018, \mn@doi
  [Monthly Notices of the Royal Astronomical Society] {10.1093/mnras/sty2261},
  481, 819–834

\bibitem[\protect\citeauthoryear{Gilman, Birrer, Treu, Nierenberg  \&
  Benson}{Gilman et~al.}{2019a}]{Gilman_2019a}
Gilman D.,  Birrer S.,  Treu T.,  Nierenberg A.,   Benson A.,  2019a, \mn@doi
  [Monthly Notices of the Royal Astronomical Society] {10.1093/mnras/stz1593},
  487, 5721–5738

\bibitem[\protect\citeauthoryear{Gilman, Birrer, Nierenberg, Treu, Du  \&
  Benson}{Gilman et~al.}{2019b}]{Gilman_2019b}
Gilman D.,  Birrer S.,  Nierenberg A.,  Treu T.,  Du X.,   Benson A.,  2019b,
  \mn@doi [Monthly Notices of the Royal Astronomical Society]
  {10.1093/mnras/stz3480}, 491, 6077–6101

\bibitem[\protect\citeauthoryear{Giocoli, Tormen, Sheth  \& van~den
  Bosch}{Giocoli et~al.}{2010}]{Giocoli_2010}
Giocoli C.,  Tormen G.,  Sheth R.~K.,   van~den Bosch F.~C.,  2010, \mn@doi
  [Monthly Notices of the Royal Astronomical Society]
  {10.1111/j.1365-2966.2010.16311.x}

\bibitem[\protect\citeauthoryear{Grazian \& Fan}{Grazian \& Fan}{2019}]{abc}
Grazian C.,  Fan Y.,  2019, A review of Approximate Bayesian Computation
  methods via density estimation: inference for simulator-models (\mn@eprint
  {arXiv} {1909.02736})

\bibitem[\protect\citeauthoryear{Harris et~al.,}{Harris et~al.}{2020}]{numpy}
Harris C.~R.,  et~al., 2020, \mn@doi [Nature] {10.1038/s41586-020-2649-2}, 585,
  357–362

\bibitem[\protect\citeauthoryear{He et~al.,}{He et~al.}{2020}]{He_2020}
He Q.,  et~al., 2020, A forward-modelling method to infer the dark matter
  particle mass from strong gravitational lenses (\mn@eprint {arXiv}
  {2010.13221})

\bibitem[\protect\citeauthoryear{He et~al.,}{He et~al.}{2021}]{He_2021}
He Q.,  et~al., 2021, Galaxy-galaxy strong lens perturbations: line-of-sight
  haloes versus lens subhaloes (\mn@eprint {arXiv} {2110.04512})

\bibitem[\protect\citeauthoryear{Hermans, Begy  \& Louppe}{Hermans
  et~al.}{2020}]{Hermans_2020}
Hermans J.,  Begy V.,   Louppe G.,  2020, Likelihood-free MCMC with Amortized
  Approximate Ratio Estimators (\mn@eprint {arXiv} {1903.04057})

\bibitem[\protect\citeauthoryear{Hermans, Delaunoy, Rozet, Wehenkel  \&
  Louppe}{Hermans et~al.}{2021}]{Hermans_2021}
Hermans J.,  Delaunoy A.,  Rozet F.,  Wehenkel A.,   Louppe G.,  2021, Averting
  A Crisis In Simulation-Based Inference (\mn@eprint {arXiv} {2110.06581})

\bibitem[\protect\citeauthoryear{Hezaveh et~al.,}{Hezaveh
  et~al.}{2016a}]{Hezaveh_2016a}
Hezaveh Y.~D.,  et~al., 2016a, \mn@doi [The Astrophysical Journal]
  {10.3847/0004-637x/823/1/37}, 823, 37

\bibitem[\protect\citeauthoryear{Hezaveh, Dalal, Holder, Kisner, Kuhlen  \&
  Levasseur}{Hezaveh et~al.}{2016b}]{Hezaveh_2016b}
Hezaveh Y.,  Dalal N.,  Holder G.,  Kisner T.,  Kuhlen M.,   Levasseur L.~P.,
  2016b, \mn@doi [Journal of Cosmology and Astroparticle Physics]
  {10.1088/1475-7516/2016/11/048}, 2016, 048–048

\bibitem[\protect\citeauthoryear{Hezaveh, Levasseur  \& Marshall}{Hezaveh
  et~al.}{2017}]{Hezaveh_2017}
Hezaveh Y.~D.,  Levasseur L.~P.,   Marshall P.~J.,  2017, \mn@doi [Nature]
  {10.1038/nature23463}, 548, 555–557

\bibitem[\protect\citeauthoryear{Hsueh, Fassnacht, Vegetti, McKean, Spingola,
  Auger, Koopmans  \& Lagattuta}{Hsueh et~al.}{2016}]{Hsueh_2016}
Hsueh J.-W.,  Fassnacht C.~D.,  Vegetti S.,  McKean J.~P.,  Spingola C.,  Auger
  M.~W.,  Koopmans L. V.~E.,   Lagattuta D.~J.,  2016, \mn@doi [Monthly Notices
  of the Royal Astronomical Society: Letters] {10.1093/mnrasl/slw146}, 463,
  L51–L55

\bibitem[\protect\citeauthoryear{Hsueh et~al.,}{Hsueh
  et~al.}{2017}]{Hsueh_2017}
Hsueh J.-W.,  et~al., 2017, \mn@doi [Monthly Notices of the Royal Astronomical
  Society] {10.1093/mnras/stx1082}, 469, 3713–3721

\bibitem[\protect\citeauthoryear{Hsueh, Enzi, Vegetti, Auger, Fassnacht,
  Despali, Koopmans  \& McKean}{Hsueh et~al.}{2019}]{Hsueh_2019}
Hsueh J.-W.,  Enzi W.,  Vegetti S.,  Auger M.~W.,  Fassnacht C.~D.,  Despali
  G.,  Koopmans L. V.~E.,   McKean J.~P.,  2019, \mn@doi [Monthly Notices of
  the Royal Astronomical Society] {10.1093/mnras/stz3177}, 492, 3047–3059

\bibitem[\protect\citeauthoryear{Hunter}{Hunter}{2007}]{matplotlib}
Hunter J.~D.,  2007, \mn@doi [Computing in Science Engineering]
  {10.1109/MCSE.2007.55}, 9, 90

\bibitem[\protect\citeauthoryear{Iqbal}{Iqbal}{2018}]{PlotNeuralNet}
Iqbal H.,  2018, HarisIqbal88/PlotNeuralNet v1.0.0,
  \mn@doi{10.5281/zenodo.2526396}, \url
  {https://doi.org/10.5281/zenodo.2526396}

\bibitem[\protect\citeauthoryear{Karchev, Coogan  \& Weniger}{Karchev
  et~al.}{2021}]{Karchev_2021}
Karchev K.,  Coogan A.,   Weniger C.,  2021, Strong-lensing source
  reconstruction with variationally optimised Gaussian processes (\mn@eprint
  {arXiv} {2105.09465})

\bibitem[\protect\citeauthoryear{Kluyver et~al.,}{Kluyver
  et~al.}{2016}]{jupyter}
Kluyver T.,  et~al., 2016, in Loizides F.,  Schmidt B.,  eds, Positioning and
  Power in Academic Publishing: Players, Agents and Agendas. pp 87 -- 90

\bibitem[\protect\citeauthoryear{Klypin, Kravtsov, Valenzuela  \& Prada}{Klypin
  et~al.}{1999}]{Klypin_1999}
Klypin A.,  Kravtsov A.~V.,  Valenzuela O.,   Prada F.,  1999, \mn@doi [The
  Astrophysical Journal] {10.1086/307643}, 522, 82–92

\bibitem[\protect\citeauthoryear{Kochanek}{Kochanek}{2004}]{Kochanek_2004}
Kochanek C.~S.,  2004, The Saas Fee Lectures on Strong Gravitational Lensing
  (\mn@eprint {arXiv} {astro-ph/0407232})

\bibitem[\protect\citeauthoryear{Koopmans}{Koopmans}{2005}]{Koopmans_2005}
Koopmans L.,  2005, \mn@doi [EAS Publications Series] {10.1051/eas:2006064},
  20, 161–166

\bibitem[\protect\citeauthoryear{Koopmans, Browne  \& Jackson}{Koopmans
  et~al.}{2004}]{SKA}
Koopmans L.,  Browne I.,   Jackson N.,  2004, \mn@doi [New Astronomy Reviews]
  {10.1016/j.newar.2004.09.047}, 48, 1085–1094

\bibitem[\protect\citeauthoryear{Laureijs et~al.,}{Laureijs
  et~al.}{2011}]{Euclid_2011}
Laureijs R.,  et~al., 2011, Euclid Definition Study Report (\mn@eprint {arXiv}
  {1110.3193})

\bibitem[\protect\citeauthoryear{Lovell}{Lovell}{2020}]{Lovell_2020}
Lovell M.~R.,  2020, \mn@doi [The Astrophysical Journal]
  {10.3847/1538-4357/ab982a}, 897, 147

\bibitem[\protect\citeauthoryear{Lovell, Frenk, Eke, Jenkins, Gao  \&
  Theuns}{Lovell et~al.}{2014}]{Lovell_2014}
Lovell M.~R.,  Frenk C.~S.,  Eke V.~R.,  Jenkins A.,  Gao L.,   Theuns T.,
  2014, \mn@doi [Monthly Notices of the Royal Astronomical Society]
  {10.1093/mnras/stt2431}, 439, 300–317

\bibitem[\protect\citeauthoryear{Mao \& Schneider}{Mao \&
  Schneider}{1998}]{Mao_1998}
Mao S.,  Schneider P.,  1998, \mn@doi [Monthly Notices of the Royal
  Astronomical Society] {10.1046/j.1365-8711.1998.01319.x}, 295, 587–594

\bibitem[\protect\citeauthoryear{McKean et~al.,}{McKean
  et~al.}{2015}]{McKean_2015}
McKean J.~P.,  et~al., 2015, Strong gravitational lensing with the SKA
  (\mn@eprint {arXiv} {1502.03362})

\bibitem[\protect\citeauthoryear{Meneghetti}{Meneghetti}{2016}]{Meneghetti_2016}
Meneghetti M.,  2016, Introduction to Gravitational Lensing

\bibitem[\protect\citeauthoryear{Miller, Cole, Louppe  \& Weniger}{Miller
  et~al.}{2020}]{Miller_2020}
Miller B.~K.,  Cole A.,  Louppe G.,   Weniger C.,  2020, Simulation-efficient
  marginal posterior estimation with swyft: stop wasting your precious time
  (\mn@eprint {arXiv} {2011.13951})

\bibitem[\protect\citeauthoryear{Miller, Cole, Forrè, Louppe  \&
  Weniger}{Miller et~al.}{2021}]{Miller_2021}
Miller B.~K.,  Cole A.,  Forrè P.,  Louppe G.,   Weniger C.,  2021, Truncated
  Marginal Neural Ratio Estimation (\mn@eprint {arXiv} {2107.01214})

\bibitem[\protect\citeauthoryear{Moore, Ghigna, Governato, Lake, Quinn, Stadel
  \& Tozzi}{Moore et~al.}{1999}]{Moore_1999}
Moore B.,  Ghigna S.,  Governato F.,  Lake G.,  Quinn T.,  Stadel J.,   Tozzi
  P.,  1999, \mn@doi [The Astrophysical Journal] {10.1086/312287}, 524, L19

\bibitem[\protect\citeauthoryear{Morningstar et~al.,}{Morningstar
  et~al.}{2019}]{Morningstar_2019}
Morningstar W.~R.,  et~al., 2019, \mn@doi [The Astrophysical Journal]
  {10.3847/1538-4357/ab35d7}, 883, 14

\bibitem[\protect\citeauthoryear{Navarro, Frenk  \& White}{Navarro
  et~al.}{1997}]{nfw}
Navarro J.~F.,  Frenk C.~S.,   White S. D.~M.,  1997, \mn@doi [The
  Astrophysical Journal] {10.1086/304888}, 490, 493–508

\bibitem[\protect\citeauthoryear{Nierenberg, Treu, Wright, Fassnacht  \&
  Auger}{Nierenberg et~al.}{2014}]{Nierenberg_2014}
Nierenberg A.~M.,  Treu T.,  Wright S.~A.,  Fassnacht C.~D.,   Auger M.~W.,
  2014, \mn@doi [Monthly Notices of the Royal Astronomical Society]
  {10.1093/mnras/stu862}, 442, 2434–2445

\bibitem[\protect\citeauthoryear{Nierenberg et~al.,}{Nierenberg
  et~al.}{2017}]{Nierenberg_2017}
Nierenberg A.~M.,  et~al., 2017, \mn@doi [Monthly Notices of the Royal
  Astronomical Society] {10.1093/mnras/stx1400}, 471, 2224–2236

\bibitem[\protect\citeauthoryear{Paszke et~al.,}{Paszke et~al.}{2019}]{pytorch}
Paszke A.,  et~al., 2019, in Wallach H.,  Larochelle H.,  Beygelzimer A.,
  d\textquotesingle Alch\'{e}-Buc F.,  Fox E.,   Garnett R.,  eds, , Advances
  in Neural Information Processing Systems 32.
Curran Associates, Inc., pp 8024--8035, \url
  {http://papers.neurips.cc/paper/9015-pytorch-an-imperative-style-high-performance-deep-learning-library.pdf}

\bibitem[\protect\citeauthoryear{{Peebles}}{{Peebles}}{1982}]{Peebles_1982}
{Peebles} P.~J.~E.,  1982, \mn@doi [\apjl] {10.1086/183911}, \href
  {https://ui.adsabs.harvard.edu/abs/1982ApJ...263L...1P} {263, L1}

\bibitem[\protect\citeauthoryear{Perreault~Levasseur, Hezaveh  \&
  Wechsler}{Perreault~Levasseur et~al.}{2017}]{PerreaultLevasseur_2017}
Perreault~Levasseur L.,  Hezaveh Y.~D.,   Wechsler R.~H.,  2017, \mn@doi [The
  Astrophysical Journal] {10.3847/2041-8213/aa9704}, 850, L7

\bibitem[\protect\citeauthoryear{Refregier, Amara, Kitching, Rassat, Scaramella
   \& Weller}{Refregier et~al.}{2010}]{Euclid_2010}
Refregier A.,  Amara A.,  Kitching T.~D.,  Rassat A.,  Scaramella R.,   Weller
  J.,  2010, Euclid Imaging Consortium Science Book (\mn@eprint {arXiv}
  {1001.0061})

\bibitem[\protect\citeauthoryear{Richings, Frenk, Jenkins, Robertson  \&
  Schaller}{Richings et~al.}{2021}]{Richings_2020}
Richings J.,  Frenk C.,  Jenkins A.,  Robertson A.,   Schaller M.,  2021,
  \mn@doi [Monthly Notices of the Royal Astronomical Society]
  {10.1093/mnras/staa4013}, 501, 4657–4668

\bibitem[\protect\citeauthoryear{Ritondale, Vegetti, Despali, Auger, Koopmans
  \& McKean}{Ritondale et~al.}{2019}]{Ritondale_2019}
Ritondale E.,  Vegetti S.,  Despali G.,  Auger M.~W.,  Koopmans L. V.~E.,
  McKean J.~P.,  2019, \mn@doi [Monthly Notices of the Royal Astronomical
  Society] {10.1093/mnras/stz464}, 485, 2179–2193

\bibitem[\protect\citeauthoryear{{Rubin}, {Ford}  \& {Thonnard}}{{Rubin}
  et~al.}{1980}]{Rubin_1980}
{Rubin} V.~C.,  {Ford} W.~K. J.,   {Thonnard} N.,  1980, \mn@doi [\apj]
  {10.1086/158003}, \href
  {https://ui.adsabs.harvard.edu/abs/1980ApJ...238..471R} {238, 471}

\bibitem[\protect\citeauthoryear{Salucci}{Salucci}{2019}]{Salucci_2019}
Salucci P.,  2019, \mn@doi [The Astronomy and Astrophysics Review]
  {10.1007/s00159-018-0113-1}, 27

\bibitem[\protect\citeauthoryear{Schneider, Smith, Macciò  \& Moore}{Schneider
  et~al.}{2012}]{Schneider_2012}
Schneider A.,  Smith R.~E.,  Macciò A.~V.,   Moore B.,  2012, \mn@doi [Monthly
  Notices of the Royal Astronomical Society]
  {10.1111/j.1365-2966.2012.21252.x}, 424, 684–698

\bibitem[\protect\citeauthoryear{{S{\'e}rsic}}{{S{\'e}rsic}}{1963}]{Sersic_1963}
{S{\'e}rsic} J.~L.,  1963, Boletin de la Asociacion Argentina de Astronomia La
  Plata Argentina, \href
  {https://ui.adsabs.harvard.edu/abs/1963BAAA....6...41S} {6, 41}

\bibitem[\protect\citeauthoryear{Simon et~al.,}{Simon et~al.}{2019}]{ELT}
Simon J.~D.,  et~al., 2019, Testing the Nature of Dark Matter with Extremely
  Large Telescopes (\mn@eprint {arXiv} {1903.04742})

\bibitem[\protect\citeauthoryear{Springel et~al.,}{Springel
  et~al.}{2008}]{Springel_2008}
Springel V.,  et~al., 2008, \mn@doi [Monthly Notices of the Royal Astronomical
  Society] {10.1111/j.1365-2966.2008.14066.x}, 391, 1685–1711

\bibitem[\protect\citeauthoryear{Suyu, Marshall, Hobson  \& Blandford}{Suyu
  et~al.}{2006}]{Suyu_2006}
Suyu S.~H.,  Marshall P.~J.,  Hobson M.~P.,   Blandford R.~D.,  2006, \mn@doi
  [Monthly Notices of the Royal Astronomical Society]
  {10.1111/j.1365-2966.2006.10733.x}, 371, 983

\bibitem[\protect\citeauthoryear{Suyu, Marshall, Blandford, Fassnacht,
  Koopmans, McKean  \& Treu}{Suyu et~al.}{2009}]{Suyu_2009}
Suyu S.~H.,  Marshall P.~J.,  Blandford R.~D.,  Fassnacht C.~D.,  Koopmans L.
  V.~E.,  McKean J.~P.,   Treu T.,  2009, \mn@doi [The Astrophysical Journal]
  {10.1088/0004-637x/691/1/277}, 691, 277–298

\bibitem[\protect\citeauthoryear{Taylor, Dye, Broadhurst, Benitez  \& van
  Kampen}{Taylor et~al.}{1998}]{Taylor_1998}
Taylor A.~N.,  Dye S.,  Broadhurst T.~J.,  Benitez N.,   van Kampen E.,  1998,
  \mn@doi [The Astrophysical Journal] {10.1086/305827}, 501, 539–553

\bibitem[\protect\citeauthoryear{Tinker, Kravtsov, Klypin, Abazajian, Warren,
  Yepes, Gottlöber  \& Holz}{Tinker et~al.}{2008}]{Tinker_2008}
Tinker J.,  Kravtsov A.~V.,  Klypin A.,  Abazajian K.,  Warren M.,  Yepes G.,
  Gottlöber S.,   Holz D.~E.,  2008, \mn@doi [The Astrophysical Journal]
  {10.1086/591439}, 688, 709–728

\bibitem[\protect\citeauthoryear{Van~Rossum \& Drake~Jr}{Van~Rossum \&
  Drake~Jr}{1995}]{python}
Van~Rossum G.,  Drake~Jr F.~L.,  1995, Python reference manual.
Centrum voor Wiskunde en Informatica Amsterdam

\bibitem[\protect\citeauthoryear{Vegetti \& Koopmans}{Vegetti \&
  Koopmans}{2009a}]{Vegetti_2009a}
Vegetti S.,  Koopmans L. V.~E.,  2009a, \mn@doi [Monthly Notices of the Royal
  Astronomical Society] {10.1111/j.1365-2966.2008.14005.x}, 392, 945–963

\bibitem[\protect\citeauthoryear{Vegetti \& Koopmans}{Vegetti \&
  Koopmans}{2009b}]{Vegetti_2009b}
Vegetti S.,  Koopmans L. V.~E.,  2009b, \mn@doi [Monthly Notices of the Royal
  Astronomical Society] {10.1111/j.1365-2966.2009.15559.x}, 400, 1583–1592

\bibitem[\protect\citeauthoryear{Vegetti, Czoske  \& Koopmans}{Vegetti
  et~al.}{2010a}]{Vegetti_2010a}
Vegetti S.,  Czoske O.,   Koopmans L. V.~E.,  2010a, \mn@doi [Monthly Notices
  of the Royal Astronomical Society] {10.1111/j.1365-2966.2010.16952.x}, 407,
  225–231

\bibitem[\protect\citeauthoryear{Vegetti, Koopmans, Bolton, Treu  \&
  Gavazzi}{Vegetti et~al.}{2010b}]{Vegetti_2010b}
Vegetti S.,  Koopmans L. V.~E.,  Bolton A.,  Treu T.,   Gavazzi R.,  2010b,
  \mn@doi [Monthly Notices of the Royal Astronomical Society]
  {10.1111/j.1365-2966.2010.16865.x}, 408, 1969–1981

\bibitem[\protect\citeauthoryear{Vegetti, Lagattuta, McKean, Auger, Fassnacht
  \& Koopmans}{Vegetti et~al.}{2012}]{Vegetti_2012}
Vegetti S.,  Lagattuta D.~J.,  McKean J.~P.,  Auger M.~W.,  Fassnacht C.~D.,
  Koopmans L. V.~E.,  2012, \mn@doi [Nature] {10.1038/nature10669}, 481,
  341–343

\bibitem[\protect\citeauthoryear{Vegetti, Koopmans, Auger, Treu  \&
  Bolton}{Vegetti et~al.}{2014}]{Vegetti_2014}
Vegetti S.,  Koopmans L. V.~E.,  Auger M.~W.,  Treu T.,   Bolton A.~S.,  2014,
  \mn@doi [Monthly Notices of the Royal Astronomical Society]
  {10.1093/mnras/stu943}, 442, 2017–2035

\bibitem[\protect\citeauthoryear{Vegetti, Despali, Lovell  \& Enzi}{Vegetti
  et~al.}{2018}]{Vegetti_2018}
Vegetti S.,  Despali G.,  Lovell M.~R.,   Enzi W.,  2018, \mn@doi [Monthly
  Notices of the Royal Astronomical Society] {10.1093/mnras/sty2393}, 481,
  3661–3669

\bibitem[\protect\citeauthoryear{Vernardos, Tsagkatakis  \& Pantazis}{Vernardos
  et~al.}{2020}]{Vernardos_2020}
Vernardos G.,  Tsagkatakis G.,   Pantazis Y.,  2020, \mn@doi [Monthly Notices
  of the Royal Astronomical Society] {10.1093/mnras/staa3201}, 499, 5641

\bibitem[\protect\citeauthoryear{Virtanen et~al.,}{Virtanen
  et~al.}{2020}]{scipy}
Virtanen P.,  et~al., 2020, \mn@doi [Nature Methods]
  {10.1038/s41592-019-0686-2}, 17, 261–272

\bibitem[\protect\citeauthoryear{Wagner-Carena, Aalbers, Birrer, Nadler,
  Darragh-Ford, Marshall  \& Wechsler}{Wagner-Carena
  et~al.}{2022}]{Wagner-Carena_2022}
Wagner-Carena S.,  Aalbers J.,  Birrer S.,  Nadler E.~O.,  Darragh-Ford E.,
  Marshall P.~J.,   Wechsler R.~H.,  2022, From Images to Dark Matter:
  End-To-End Inference of Substructure From Hundreds of Strong Gravitational
  Lenses (\mn@eprint {arXiv} {2203.00690})

\bibitem[\protect\citeauthoryear{Waskom}{Waskom}{2021}]{seaborn}
Waskom M.~L.,  2021, \mn@doi [Journal of Open Source Software]
  {10.21105/joss.03021}, 6, 3021

\bibitem[\protect\citeauthoryear{Zwicky}{Zwicky}{1933}]{Zwicky_1933}
Zwicky F.,  1933, \mn@doi [Helv. Phys. Acta] {10.1007/s10714-008-0707-4}, 6,
  110

\makeatother
\end{thebibliography}

\appendix

\section{Sérsic source}
\label{apx:sersic}

Here we describe the elliptical coordinates $(r_x,r_y)$, the normalization $k_n$ and the index $n$ that enter in the modeling of the Sérsic profile in \autoref{eq:sersic}.

The transformation from Cartesian $(x,y)$ to elliptical coordinates $(r_x,r_y)$ is given by
\begin{equation}
    \begin{pmatrix} 
        r_x \\ r_y 
    \end{pmatrix} = 
    \begin{pmatrix} 
      \sqrt{q} & 0 \\ 0 & 1/\sqrt{q} 
    \end{pmatrix} 
    \begin{pmatrix} 
        \cos\phi_s & \sin\phi_s \\ -\sin\phi_s & \cos\phi_s 
    \end{pmatrix} 
    \begin{pmatrix} 
        x - x_0 \\ y - y_0 
    \end{pmatrix},
\end{equation}
where $\phi_s$ is the rotation angle, $q_s$ is the axis ratio, and $(x_0, y_0)$ is the center of light position. 

The normalization $k_n$ is related to the index $n$ by an implicit transcendental equation in terms of the complete and lower incomplete gamma functions $2\gamma(2n, k_n)=\Gamma(2n)$. We use the expansion in series from \cite{Ciotti_1999}, valid over a wide range of indices $n$, stopping at order $\order{n^{-3}}$.

\section{Line-of-sight halos as effective subhalos}
\label{apx:los}

Following \cite{Sengul_2020}, \gls*{los} halos at comoving distance $\chi$ can be treated as subhalos on the main-lens plane with an effective projected mass density given by:
\begin{equation}
    \Sigma_{\chi, \mathrm{eff}}(D_l\vec{x};m_{200},r_s,\tau)= \Sigma(D_l\vec{x};m_{200,\mathrm{eff}},r_{s,\mathrm{eff}},\tau).
\end{equation}
The effective scale radius $r_{s,\mathrm{eff}}$ and mass $m_{200,\mathrm{eff}}$ are respectively
\begin{equation}
    r_{s,\mathrm{eff}}=\cfrac{D_l}{g(\chi)D_\chi}r_s,
\end{equation}
and
\begin{equation}
    m_{200,\mathrm{eff}}=f(\chi)\cfrac{\Sigma_{\mathrm{cr}, l}}{\Sigma_{\mathrm{cr}, \chi}}\left(\cfrac{D_l}{g(\chi)D_\chi}\right)^2m_{200}.
\end{equation}
The piecewise functions $f(\chi)$ and $g(\chi)$ are:
\begin{equation}
    f(\chi)= \begin{cases}
            1-\beta_{\chi l} & \chi \leq \chi_l \\
            1-\beta_{l\chi}  & \chi > \chi_l \\
            \end{cases},
\end{equation}
and
\begin{equation}
    g(\chi)= \begin{cases}
            1 & \chi \leq \chi_l \\
            1-\beta_{l\chi}  & \chi > \chi_l \\
            \end{cases},
\end{equation}
with $\beta_{ij}=\cfrac{D_{ij}D_s}{D_jD_{is}}$, where $D_i$ is the angular diameter distance from the observer to plane i, and $D_{ij}$ is the angular diameter distance from lens plane i to lens plane j, and $\chi_l$ is the comoving distance to the main-lens plane.
We have also introduced the critical surface density at plane i
\begin{equation}
    \Sigma_{\mathrm{cr}, i}\equiv \cfrac{c^2D_s}{4\pi G D_i D_{is}}.
\end{equation}

\bsp	
\label{lastpage}

\end{document}